\documentclass[12pt,preprint]{aastex}

\newcommand{\ud}{\textrm{d}}

\newcommand{\LA}{\ensuremath{\textrm{Ly}\alpha}}
\newcommand{\HA}{\ensuremath{\textrm{H}\alpha}}
\newcommand{\HB}{\ensuremath{\textrm{H}\beta}}

\newcommand{\OII}{\ensuremath{\textrm{[O\sc{ii}]}\lambda\textrm{3727}}}
\newcommand{\OIIcon}{\ensuremath{\textrm{[O\sc{ii}]}\lambda\textrm{3727}}}

\newcommand{\HeII}{\ensuremath{\textrm{He\sc{ii}}}}

\newcommand{\OIIIcon}{\ensuremath{\textrm{[O\sc{iii}]}\lambda\lambda\textrm{4959,5007}}}
\newcommand{\OIII}{\ensuremath{\textrm{[O\sc{iii}]}\lambda\lambda\textrm{4959,5007}}}

\newcommand{\OIIIa}{\ensuremath{\textrm{[O\sc{iii}]}\lambda\textrm{4959}}}
\newcommand{\OIIIb}{\ensuremath{\textrm{[O\sc{iii}]}\lambda\textrm{5007}}}
\newcommand{\OIIIsin}{\ensuremath{\textrm{[O\sc{iii}]}}}
\newcommand{\NII}{\ensuremath{\textrm{[N\sc{ii}]}\lambda\lambda\textrm{6548,6584}}}
\newcommand{\NIIa}{\ensuremath{\textrm{[N\sc{ii}]}\lambda\textrm{6548}}}
\newcommand{\NIIb}{\ensuremath{\textrm{[N\sc{ii}]}\lambda\textrm{6584}}}
\newcommand{\NIIsin}{\ensuremath{\textrm{[N\sc{ii}]}}}

\newcommand{\broadband}{\ensuremath{\mathrm{B}}}
\newcommand{\narrowband}{\ensuremath{\mathrm{N}}}

\newcommand{\narrowbandL}{\ensuremath{\mathrm{N}}}
\newcommand{\meanf}[1]{\ensuremath{\bar{f}_{\mathrm{#1}}}}
\newcommand{\fluxl}{\ensuremath{f_{\mathrm{l}}}}
\newcommand{\ew}{\ensuremath{W}}
\newcommand{\fluxc}{\ensuremath{f_{\mathrm{c}}}}

\slugcomment{Accepted for publication in PASP}

\shorttitle{Selection of ELGs using narrow-band filters in optical windows}

\begin{document}


\title{A contribution to the selection of emission-line galaxies using narrow-band filters in the optical airglow windows}


\author{S. Pascual and J. Gallego and J. Zamorano}
\affil{Departamento de Astrof\'{\i}sica y Ciencias de la Atm\'osfera,
    Facultad de C.C. F\'{\i}sicas,
    Universidad Complutense de Madrid, E-28040, Madrid, Spain}
\email{spr@astrax.fis.ucm.es}



\begin{abstract}
  Emission line galaxies are an invaluable tool for our understanding
  of the evolution of galaxies in the Universe. Imaging of deep fields
  with narrow-band filters allows not only the selection of these
  objects, but also to infer the line flux and the equivalent width of
  the emission line with some assumptions.  The narrow-band filter
  technique provides homogeneous samples of galaxies in small comoving
  volumes in the sky.  We present an analysis of the selection of
  emission-line galaxies using narrow-band filters.  Different methods
  of observation are considered: broad-band -- narrow-band filters
  and two broad-band and one narrow-band
  filters.  We study also the effect of several lines entering
  simultaneously inside the filters (this is the case of \HA).  In
  each case the equations to obtain the equivalent width and line flux
  from the photometry are obtained.  Candidates to emission-line
  objects are selected by their color excess in a magnitude-color
  diagram.  For different narrow-band filters, we compute the mean
  colors of stars and galaxies, showing that, apart from galaxies,
  some types of stars could be selected with certain filter sets.  We
  show how to compute the standard deviation of the colors of the
  objects even in the usual case when there are not enough objects to
  determine the standard deviation from the data. We present also
  helpful equations to compute the narrow-band and the broad-band
  exposure times in order to obtain minimum dispersion in the ratio of
  fluxes of both bands with minimum total exposure time.
\end{abstract}






\section{Introduction}

Emission line galaxies (hereafter ELGs) are an invaluable tool for our
understanding of the evolution of galaxies in the Universe.  Faint
galaxies can be hard to confirm spectroscopically, while the ELGs are
generally easy to identify. Furthermore, the emission lines are
produced within regions related either with star formation or with the
active galactic nuclei (AGN) phenomena.

Before the advent of CCD detectors, surveys for emission-line galaxies
used photographic plates with objective prisms in Schmidt telescopes
\citep[e.g. ][]{1975ApJ...202..591S,1976ApJS...32..217S,1977ApJS...34...95M,
1983ApJ...272...68W,1983ApJS...51..171P,
1989ApJS...70..447S,1994ApJS...95..387Z,1996ApJS..105..343Z}.

Several low redshift surveys have made use of the objective-prism
technique with CCD detectors to circumvent the limit on wavelength
and quantum efficiency imposed by photographic plates
\citep[e.g.,][]{2001ApJ...548..585S,2004AJ....127.1943G,2005MNRAS.359..930B}.
However the objective prism, as any system of slitless spectroscopy,
blends the flux of the object with that of the dispersed sky,
reducing the sensitivity of the survey for the objects with less
contrast between the continuum and the line (that is, with small
equivalent width). This effect, together with relatively small
aperture of the telescopes used, limits these surveys to detect
bright sources in a near volume of the space.
    
A more efficient approach to detect ELGs is to use narrow-band imaging
\citep{1985ApJ...299L...1D,1991ApJ...377L..73L,1993ApJ...404..511M,1995AJ....110..963T}.
This technique reduces significantly the contribution of the sky
brightness, as it is admitted in a small range of wavelengths. The sky
background, that is the most significant limitation in the detection
of objects in deep broad band images, is greatly reduced.  A small
wavelength range increases also the contrast between the emission line
and the continuum.  Narrow-band imaging has been used also to detect
both galactic \citep{1999A&AS..135..487R} as well as extragalactic
\citep{2002PASJ...54..883O,2003A&A...405..803C,2003AJ....125..514A}
planetary nebulae.

In order to use the narrow pass-band filters with maximum efficiency,
regions of the night sky spectrum with a minimum background are
selected.  In the optical wavelengths, the windows in the Meinel OH
bands around 8200\AA{} and 9200\AA{} have been used to center
narrow-band filters.  Most of the interest has gone to the region
around 8200\AA{}, where the CCD detectors still have a good quantum
efficiency.

Similar to the narrow-band imaging is the use of a tunable filter
\citep[TF, ][]{1998PASA...15...44B}.  A tunable filter works like a
Fabry-Perot interferometer modified to produce a wide region of
interference and to cover a range of wavelengths.

Compared with narrow-band imaging, a TF provides a pseudo-spectrum,
when the spectral range is explored with small wavelength steps. As a
disadvantage, the TF requires a dedicated instrument, whereas
narrow-band imaging demands only the narrow-band filter and can be
carried out in any imaging facility.  Both narrow-band imaging and TFs
need very long exposure times to achieve the necessary signal-noise
ratio.

The search for emission-line galaxies using a given narrow-band
filter is open to different lines at different redshifts.  Lines
whose redshifted wavelength fall inside the narrow-band filter
range could be detected. When the emission-line is bright enough
the galaxy will be selected. The more prominent 
lines that can be detected in the 8200\AA{} and 9200\AA{} 
atmospheric windows are {\HA $\lambda6563$} (at redshift $z\simeq0.24$ 
and $z\simeq0.4$), {\OIIIcon} ($z\simeq0.6$ 
and $z\simeq0.8$), {\OIIcon} ($z\simeq1.2$ and $z\simeq1.5$) and {\LA $\lambda1216$}  
($z\simeq5.7$ and $z\simeq6.5$).

The \HA{} line is an excellent tracer of the Star Formation Rate
\citep[SFR, ][]{1998ARA&A..36..189K, 2001MNRAS.323..887C}, provided
that the ionizing flux comes from young stars and not from non-thermal
activity.  Other commonly used SFR tracers are the far infrared (FIR)
and the UV.  The three tracers share sensitivity to the parameters of
the Initial Mass Function.  \HA{} SFR is affected by obscuration but
is not very sensitive to metallicity.  The FIR is not affected by
obscuration, but there is uncertainty in how the total (8-1000$\mu$m)
luminosity is computed from monochromatic measurements (e.g. Spitzer's
24$\mu$m) and the contribution of different non-SFR related components
(cirrus, hot dust) to the FIR. Finally the UV is heavily affected by
obscuration.  As shown by \citet{2003apj...586..794b}, obscuration
corrected \HA{} is consistent, to a factor of 2, with the
\emph{summed} SFRs estimated using UV and FIR(8-1000$\mu$m).
Consequently, \HA{} observations of galaxy samples with UV and FIR
data provides an invaluable tool to understand the evolution of the
SFR and the role of obscuration in the determination of global SFR for
galaxies.  Some works have used narrow-band imaging to select \HA{}
candidates in the 8200\AA{} window: \citet{2001A&A...379..798P,
2003ApJ...586L.115F, 2004ApJ...601..805U} and
\citet{2006PASJ...58..113A}.

Given that small amounts of dust reddening can completely extinguish
\LA-emission, \HA{} is the strongest emission line in many starburst
galaxies.  The \OIII{} doublet is in many cases the other strongest
emission line. In fact, the \OIII/\HA{} ratio is 0.6 in local
star-forming galaxies
\citep{2005MNRAS.362.1143M}.  This line is
not used to trace the star formation but to trace the power of AGN
\citep[see, for example][]{2003MNRAS.346.1055K}. Thus, \OIII{}
emitters can be a tool to unveil AGN at different redshifts.

\OII{} is also intense when compared with \HA{}, with a ratio
\OII/\HA{} = 0.45 \citep{1992ApJ...388..310K}. However, the ratio
depends on luminosity and metallicity \citep{2001ApJ...551..825J} and
varies from sample to sample.  This line is used as a star-formation
indicator at redshifts where \HA{} is outside the visible range (which
happens at $z > 0.4$). SFR based on \OII{} comes from the fact that
there is a good correlation between \OII{} emission and \HA{}
emission. \OII{} is used as a proxy for the \HA{} emission. Different
calibrations of SFR based on \HA{} are used \citep[see,
e.g. ][]{1989AJ.....97..700G,1998ARA&A..36..189K,2002MNRAS.332..283R,2004AJ....127.2002K}
but, in general, the correlation between SFR and \OII{} is affected
intensely by extinction and metallicity \citep{2006MNRAS.369..891M}.

In the assumption of case B recombination, two-thirds of the Lyman
continuum photons are reprocessed as \LA{}. Therefor, young starbursts
should be easily detected by their \LA{} emission. However,
observations of nearby starbursts \citep[e.g.][]{1988ApJ...326..101H,
1992ApJ...399L..39C} revealed in most starburst galaxies a much weaker
\LA{} than predicted by simple models of galaxy formation.
To explain this low flux, \citet{1993ApJ...415..580C} suggested that
\LA{} strong emission requires either a low abundance of dust or an
AGN. However, \citet{1996ApJ...470..189G} demonstrated that pure
extinction could not explain the weak \LA{} fluxes. Results suggest
that \LA{} photons are decoupled from the continuum radiation
resulting in a increased sensitivity to dust \citet{1990ApJ...350..216N}.

\LA{} systems at $z>5$ are a very useful probe of
the properties of very young star-forming systems at the epoch after
reionization. Unfortunately, the data obtained to the date
\citep[e.g. ][]{2003PASJ...55L..17K,2004AJ....127..563H,
2004ApJ...611...59R, 2005A&A...430L..21W, 2005PASJ...57..165T,
2006ApJ...638..596A,2006PASJ...58..313S} are not sufficient to place
strong constraints on the fundamental properties.  The \LA{} line is
not used commonly as a SFR tracer.  Resonant scattering of \LA{}
photons by neutral atomic hydrogen affects their relation with the SFR
in a galaxy \citep{1993ApJ...415..580C}.

Narrow-band imaging produces approximately volume-limited samples,
since the narrow-observed bands correspond to small windows in
redshift space. The objects are selected with a well defined limit in
equivalent width and the line flux can be transformed into luminosity
with some simple assumptions.  Narrow-band imaging thus provide line
luminosities for a volume-limited sample of emission line galaxies.
In the case of detecting lines used as star formation tracers, the
sample would be directly SFR-selected, except for the AGN
contribution.  The problem with this approach is that stars
contaminating the sample or contributions from different emission
lines cannot be separated only with narrow-band imaging. Additional
assumptions \citep[on the luminosity functions of ELGs,
e.g. ][]{2001ApJ...550..593J} or additional data \citep[see, e.g., the
color-color diagrams of][]{2003ApJ...586L.115F} are needed to complete
the scenario.

The aim of this work is to contribute to several aspects of the
narrow-band imaging in order to optimize the results of the
observations.  Section~\ref{sec:description} describes the
characterization of the filters used through this work.

In Section \ref{sec:contamination} we deal with the different sources
of contamination, either galactic or extragalactic that can be found
in an narrow-band survey and different methods used to detect them.
Section~\ref{sec:el} describes how to infer the line and continuum
fluxes (and the corresponding equivalent width) from the measured
magnitudes in a variety of scenarios, including more than one
broad-band filter and more than one line inside the narrow-band
filter. We study the method to select the candidates, based solely on
broad and narrow-band colors in Sect.~\ref{sec:dispq}.  We study in
detail also the comoving volume covered by a narrow-band selected
sample. Numeric examples with three filter set, including broad and
narrow-band filters, are shown in Section~\ref{sec:examples}. Although
we have based our examples in two surveys carried out using filters at
the 8200\AA{} and 9200\AA{} windows, the analysis and methods
described in this work are valid for a wide variety of narrow-band
imaging surveys, including those extended into the IR. Finally, in
Sect.~\ref{sec:optimal}, we present helpful equations to compute the
narrow-band and the broad-band exposure times in order to obtain
minimum dispersion in the ratio of fluxes of both bands with minimum
total exposure time.  Conclusions are exposed in
Sect.~\ref{sec:conclusiones}.

\section{Characterization of the filters}
\label{sec:description}

In general, the filters used in CCD photometry are characterized by
their transmittance curve $T(\lambda)$. The fiducial transmittances are
usually measured by the observatories with collimated beams and at
room temperature. The total transmittance includes the contribution of
the quantum efficiency (QE) of the detector and the transmittances of
the optical systems in the light path.  The working temperature,
usually lower than the room temperature, produces a blue shift of the
transmittance of the narrow band filters. A converging light beam
passing through the filter produces a second blue shift, as well as a
decrease of the overall transmittance and a broadening of the
transmittance profile. These effects depend on the focal ratio of the
beam passing through the filter.

Except in Section \ref{sec:dispq}, where absolute source flux are
needed for statistics, we use always the transmittance normalized to
peak unity ($\mathcal{T}$) instead of the absolute transmittance
($T$). Both are related by $T = T_{\mathrm{peak}} \mathcal{T}$, where
$T_{\mathrm{peak}}$ is the maximum value of the transmittance.

The mean wavelength of the band, $\lambda_0$, is computed 
in agreement with the conventional definition 
\citep[see for example][]{1995pasp..107..945f, 2003A&A...401..781F}. 
\begin{equation}
  \lambda_0 \equiv \frac{\int \lambda\, \mathcal{T}(\lambda)\,\ud
    \lambda}{\int \mathcal{T}(\lambda)\, \ud \lambda}
\end{equation}
$\Delta$ is the width of the filter computed from $\mathcal{T}$. It
agrees with the definition of the equivalent width of the band
transmittance profile $W_0$ of \citet{2003A&A...401..781F}:
\begin{equation}
  W_0 \equiv \Delta \equiv \int \mathcal{T}(\lambda)\,\ud \lambda
\end{equation}
We use $\Delta$ instead of $W_0$ and reserve $W$ for the equivalent
width of the emission lines.
  
$\mu^2$  \citep[Eq. 7 of ][]{2003A&A...401..781F} is the second-order 
central momentum of the transmittance profile:
\begin{equation}
  \mu^2 \equiv \frac{\int (\lambda - \lambda_0)^2\,
    \mathcal{T}(\lambda)\,\ud \lambda}{\int \mathcal{T}(\lambda)\,\ud
    \lambda}
  \label{eq:def:mu}
\end{equation}

The flux calibration of the narrow-band images can be achieved using
spectrophotometric standard stars, by obtaining synthetic magnitudes
using the known SED of the spectrophotometric stars.

There are different choices for the election of the zero point of the
narrow-band filter magnitude.  The particular election does not affect
the selection of objects or the measured fluxes.  Different zero
points in the bands only introduce an offset in the broad-narrow
color.

One approach is to use AB magnitudes \citep{1974ApJS...27...21O} or ST
magnitudes (J. Walsh 1995)\footnote{Optical and UV spectrophotometric standard stars \url{http://www.eso.org/observing/standards/spectra}} that provide a fixed physical magnitude scale. In
other approximations, a broad-band filter is chosen as reference and
the zero point of the narrow band is obtained by fixing a zero color
for a particular SED.  We adopt, unless noted otherwise, the
normalization of color zero for the flat $f_\lambda$ spectrum.  This
is equivalent to make equal the zero points of both bands.

\section{Contaminating sources}
\label{sec:contamination}

The redshift of each source entering the narrow-band filter is unknown
\emph{a priori}.  Sources with strong emission lines (particularly,
but not only, \HA{}, \OIIIcon, {\OIIcon} and \LA{}) produce a
narrow-band flux excess. Different emission lines at different
redshifts cannot be distinguished using only the narrow-band excess
emission. Additionally, other sources (either galactic or
extragalactic) without emission lines can exhibit a narrow-band flux
excess when their SEDs passes through the filter set used for ELG
selection. Galaxies and stars with abundant absorption features in the
atmospheric windows studied can potentially appear in narrow-band
surveys as emitting objects. This aspect has to be studied for each
filter set particularly (in Section~\ref{sec:examples}, examples with
different filters are shown).

In general, stars show colors similar to those of the black body of
their effective temperature. For the late spectral types, the presence
of molecular bands change the trend and large colors can be observed.
In the specific wavelength range of the two atmospheric windows, the
absorption produced by some molecular species present in the
atmosphere of cold stars (TiO, VO) dominates. For late K stars,
broad-narrow colors diverge from the black body. White dwarfs with
broad Hydrogen absorption lines would produce also a narrow-band flux
depression.

It is important to note that depending on the filters used,
particularly on the relative position of their $\lambda_0$, molecular
bands of cold stars could produce an increase of the color.  Those
stars with a large color excess produced by absorption lines can be
misclassified as ELGs. As the ratio of stars to galaxies decreases
with increasing magnitude \citep[see, for example,
][]{2003MNRAS.343.1013K}, the number of contaminating stars do not
dominate the candidate selection at faint magnitudes.  In any case,
particular care has to be taken in the design or election of the
filter set in order to avoid this source of contamination.

Some types of stellar objects show emission lines in the wavelength
interval where the filters are defined. Novae, symbiotic stars,
cataclysmic variables (in general, low mass stars with
either coronal activity or binaries) can show emission in the line
Pa9 (9229 \AA). After a nova explosion there is
emission in the multiplet O\textsc{i} near 9261-9266 \AA{} and
the line \HeII{} in 8236 \AA. Lines of N\textsc{i} and Na\textsc{ii}
between 8185 and 8242 \AA{} \citep{2003ApJ...596.1229R} can also be found.
  
Massive stars, such as some Wolf-Rayet, can exhibit emission lines
of Pa9 (9229 \AA) and in \HeII+C\textsc{iv} (8236 \AA)
\citep{2002A&A...392..653C}.

The equivalent widths of the lines in the range covered by the
narrow-band filters is of the order of tens of Angstroms. 
Although the relative frequency of appearance
of these types of stars is very low, they can be also selected by
their color excess.

\subsection{Classification schemes}

Further analysis is needed in other to classify the different objects 
selected by their narrow-band flux excess.

To separate the stars, we can apply morphological methods to the
selected candidates. A example of that approach appears in
\citet{2003A&A...410...17M}. An alternative, and commonly used,
approximation is to make use of the artificial neural network of the
source-characterization program SExtractor
\citep{1996a&as..117..393b}. Nevertheless, these classification methods
depend on the SNR and seeing of the images.  It can be difficult to
classify candidates morphologically when the overall SNR is low or the
seeing is not optimal. Multicolor photometry can be
also used to distinguish stars using a color criterion (we show a
simple example in Sect.~\ref{sec:contpoli})

Methods to classify extragalactic sources range from statistical
corrections \citep{2001ApJ...550..593J} to color-color criteria and
photometric redshift determination. For example, \citet{2001A&A...379..798P} recomputed
the statistical correction of \citet{2001ApJ...550..593J} to find that the estimated
contamination from galaxies at $z > 0.24$ was negligible.

\citep{2003ApJ...586L.115F} uses three photometric bands ($B$, $R_C$ and
$I_C$) and the GISSEL96 models \citep{2003MNRAS.344.1000B} to develop
a color criterion that allow the authors to select \HA{} emitters at
$z$=0.24.  Following a similar procedure, but using spectra of emission-line 
galaxies rather than models, \citet{2006astro.ph.10846L} uses
five photometric bands ($B$, $R_C$, $V$, $i'$ and $z'$).

\LA{} surveys use dropout technique in filters bluer than $i$ in
order to assure that the Lyman break ($\lambda$ = 912\AA) lies bluewards
the narrow-band filter.  For example, \citet{2005PASJ...57..165T}
request that the candidates are not detected in $B$, $V$ and $R$, in addition
to other criteria color criteria.

Finally, we mention a  more complex classification scheme. It is presented in
\citet{2001A&A...365..681W} and it is used in the CADIS survey
\citep{2003A&A...402...65H}. Twelve medium pass band filters are used,
in addition to broad-band $B$, $R$, $I$, $J$ and $K'$, to classify the
objects in three broad categories: stars, quasars (including AGN) and galaxies.

\section{Measured flux and equivalent width of the emission line}
\label{sec:el}
Fluxes of the objects through the filters are obtained integrating
the spectrum of the object through the filter profile.
The narrow-band filter technique allows, not only to select the
objects with a possible emission line, but also to compute the value
of the line flux and the equivalent width of the line. 

In order to obtain the line and continuum flux for the selected
objects, several simplifications, that are studied in this section,
need to be made.  First, the line profile is assumed infinitely thin
when compared with the width of the narrow-band filter.  Second, the
positioning of the emission within the transmittance profile of the
used filters is very close to the center of the narrow-band
filter. With simple assumption we can obtain a mean wavelength for the
emission line. Finally, the continuum flux can be modeled by an
analytic function, either a power-law or a polynomial. It is critical
to have a good estimate of the continuum flux.

\subsection{The infinitely thin line approximation}
\label{sec:itl}
The FWHM of the lines of star-forming galaxies is related with the
mass of the object and typically is less than $\sim$ 10 \AA.  To
estimate the influence of the finite width of the emission line we
have integrated Gaussian profiles through Gaussian and rectangular
narrow-band filters and computed the recovered line flux. The FWHM
increases with redshift a factor $1+z$, so the infinitely thin line
approximation can be valid, recovering more than 80\% of the line
flux, up to $z\sim4$ for very narrow filters ($\Delta\sim50\AA$).  For
wider filters, ($\Delta\sim150\AA$), we recover more than 80\% of the
flux up to $z\sim10$.  These limit redshifts are lower for sources with
broader lines, such as AGN.  The ELGs can still be selected using the
methods described in Sect. \ref{sec:dispq} but the fluxes of the lines
computed with this section's equations will underestimate the total
flux.

\subsection{Line flux and EW of a source with an infinitely thin line}
\label{sec:fl}
We assume that the SED of a source at redshift $z$ with a emission
line of rest wavelength $\lambda_r$ can be expressed as the sum of a
continuum contribution and an infinitely thin emission line with flux
\fluxl{}.  The emission line is represented analytically by a delta of
Dirac centered in $\lambda_z=\lambda_r(1+z)$.

In the following we assume that we have identified the emission line
detected inside the narrow-band filter. The exact wavelength of the
line is also unknown, but it is located in a very narrow range inside
the narrow-band filter. We handle the determination of an approximate
redshift for the line in Sect.~\ref{sec:meanz}.

With the assumed conditions, the resulting SED is:
\begin{equation}
  f(\lambda)=\fluxc{}(\lambda) + \fluxl\ \delta(\lambda-\lambda_z)
  \label{eq:flujodelta}
\end{equation}
with $\fluxl$ constant and $\fluxc{}(\lambda)$ an arbitrary function
of the wavelength. Note that a negative value of $\fluxl$ produces an
absorption line.

We define the mean flux through the filter as:
\begin{equation}
  \meanf{} \equiv \frac{\int f(\lambda)\,T(\lambda)\,\lambda\,\ud \lambda}{\int 
    T(\lambda)\,\lambda\,\ud \lambda} = 
  \frac{\int f(\lambda)\,\mathcal{T}(\lambda)\,\lambda\,\ud \lambda}{\int 
    \mathcal{T}(\lambda)\,\lambda\,\ud \lambda}
  \label{eq:def:meanflux}
\end{equation}
The mean flux of the SED of Eq.~\ref{eq:flujodelta} is:
\begin{equation}
  \meanf{} = \bar{\fluxc{}} + 
  \left(\frac{\mathcal{T}(\lambda_z)}{\Delta}\right)
  \left(\frac{\lambda_z}{\lambda_0}\right)\fluxl
  \label{eq:flujo1}
\end{equation}

The mean flux is a function of the redshift of the source $z$ (or a
function of $\lambda_z$, as both are related)\footnote{We will use
both variables interchangeably, as appropriated}.  The mean flux
depends on the normalized transmittance of the filter at $\lambda_z$,
the wavelength where the emission line is located and on the mean
value of the continuum flux. We have not introduced yet any constraint
to the shape of the continuum.

\subsubsection{Effective width of the filters}

Returning to Eq.~\ref{eq:flujo1}, we introduce the \emph{effective
width} of the filter $\Delta'$:
\begin{equation}
  \Delta'(\lambda_z) \equiv \left(\frac{\Delta}{\mathcal{T}(\lambda_z)}\right)
  \left(\frac{\lambda_0}{\lambda_z}\right)
  \label{eq:def:ewidth}
\end{equation}

This definition is slightly different of the width obtained
integrating the normalized transmittance. Both $\Delta$ and $\Delta'$
are independent of the maximum transmittance of the filter. $\Delta'$
depends on the redshift of the source.

The value of $\Delta'$ is close to $\Delta$ in the center of the
filter, in the region where the relative transmittance is close to
unity. In the wings of the filter, the effective width increases and
tends to infinity when the transmittance tends to zero. The physical
meaning is that a emission line does not contribute to the integrated
flux if it is positioned on a wavelength where the transmittance is
near zero. The term in Eq~\ref{eq:flujo1} containing the line flux is
effectively zero when the effective width tends to infinity.

The result of \citet{1990PASP..102.1217W} is based on square filters.
We can recover this result making the transmittance equal to unity and
making $\lambda_0 = \lambda_z$.  In that case, the effective width of
the filter reduces to:
\begin{equation}
  \Delta'(\lambda_z) = \Delta\left(\frac{\lambda_0}{\lambda_z}\right)
  \simeq \Delta
\end{equation}

In general, the flux equation \ref{eq:flujo1} can be rewritten finally
as:
\begin{equation}
  \meanf{}=\bar{\fluxc{}}+\frac{1}{\Delta'(\lambda_z)}\fluxl
  \label{eq:flujo2}
\end{equation}

\subsubsection{Estimation of mean redshift}
\label{sec:meanz}
We focus on the position of the emission line within the narrow-band
filter.  The definition of effective width depends on the unknown
redshift of the source.  As the candidates are selected by their color
excess in a narrow-band filter, the object with emission lines will be
distributed in a small range of wavelengths, given mainly by the
transmittance of the narrow band filter.  If $\ud n$ is the number of
galaxies with a emission line, whose rest wavelength is $\lambda_r$,
the number of selected objects is
$P(\lambda_z,m_{\broadband},m_{\narrowband})\,\ud n$. If we suppose
that the probability of selecting a given object can be factorized in
a term depending on $\lambda_z$ and in terms depending on the
magnitudes, the mean value of an arbitrary function $h(\lambda_z)$ can
be written as:
\begin{equation}
  \bar{h}=\frac{\int h(\lambda_z)\,p(\lambda_z) \,\ud n}{\int
  p(\lambda_z)\,\ud n}
  \label{eq:mean1}
\end{equation}

We can assume also that the probability $p(\lambda_z)$ 
is dominated by the shape of the narrow-band filter and write:
\begin{equation}
  p(\lambda_z) \propto \Delta^{'-1}_{\narrowband}
\end{equation}
Assuming a constant comoving density of galaxies inside the
narrow-band filter, the number of galaxies is proportional to the
comoving volume $V$ \citep[for the explicit expression of the comoving
volume, see][ and references therein]{1999astro.ph..5116H}. The final
equation is:
\begin{equation}
  \bar{h}=\frac{\int
      h(\lambda_z)\,\Delta^{'-1}_{\narrowband}(\lambda_z)\,\ud n}
      {\int \Delta^{'-1}_{\narrowband}(\lambda_z)\,\ud n} = \frac{\int
      h(\lambda_z)\,\lambda_z
      \mathcal{T}_{\narrowband}(\lambda_z)\,\ud V(z)} {\int \lambda_z
      \mathcal{T}_{\narrowband}(\lambda_z)\,\ud V(z)}
	   \label{eq:mean2}
\end{equation}
The mean value of $\lambda_z$ can be directly computed from last equation.
The mean values computed with this method depend slightly on the
cosmology parametrization selected (except on the value of the Hubble constant) and on
the emission line chosen. 

\subsection{Line flux and EW with different filter layouts}
The narrow-band filter layout used determines the approach used to
approximate the continuum near the emission line. In a general case, we have
$n$ filters and $n$ equations like Eq. \ref{eq:flujo1}, one for each
filter. This provides us with, at most, $n - 1$ parameters to
parametrize the continuum and one parameter to characterize the line,
($\fluxl$).

A power-law continuum ($\fluxc \propto \lambda^\beta$) can be fitted
with at least three bands. It is even possible to assume a given power-law
index (for example, the $\beta=-1$ for $\LA$ emitters in
\citet{2006ApJ...645L...9A}).  Another possible functional form for
the continuum is a polynomial of degree $n - 2$.

Most surveys use one narrow-band filter on the region of interest and
one or more broad-band filters
\citep[e.g.][]{2001A&A...379..798P,2004AJ....127..563H,2006ApJ...638..596A}. Other
surveys use more several contiguous narrow-band filters and a
medium-band filter as a veto filter \citep{2006A&A...455...61W} or
more complicated layouts \citep{2006astro.ph..9497H}.

In the following sections we study two different cases of the
polynomial functional form. With two broad-band filters, a linear
dependency of the continuum flux with wavelength can be assumed. In
the case of having one broad-band filter, the continuum has a constant
value.

\subsubsection{Three filters and a linear continuum}
\label{sec:contpoli}

We develop here a general solution for the case of three filters
covering the line. Solutions for other cases with more filters can be
easily obtained.

The continuum is approximated by a linear term with slope
$\gamma$.  For simplicity, we select the parameters so that the
linear term vanish at $\lambda_z$:
\begin{equation}
  \fluxc(\lambda) = \fluxc(\lambda_z) + \gamma (\lambda -\lambda_z)
\end{equation}

Using Eq.~\ref{eq:def:meanflux}, the effective flux is:
\begin{equation}
  \meanf{} = \fluxc{} + \gamma (\lambda_0 + \frac{\mu^2}{\lambda_0} -
  \lambda_z) + \frac{\fluxl}{\Delta'(\lambda_z)}
\end{equation}
being $\mu$ the second order central moment of the transmittance of the
filter (Eq.~\ref{eq:def:mu}). We rename the factor of $\gamma$ as $s$:
\begin{equation}
  s = \lambda_0 + \frac{\mu^2}{\lambda_0}-\lambda_z
\end{equation}
We also substitute $\lambda_z$ by $\bar\lambda_z$ in $s$ and in
$\Delta'$.  Equation~\ref{eq:flujo2} is written here for the three
filters covering the line (denoted A, B and N).
\begin{eqnarray}
  \meanf{A} & = & \fluxc{} + \gamma s_{\mathrm{A}} + {1 \over
  \Delta'_{\mathrm{A}}}\fluxl\\ \meanf{B} & = & \fluxc{} + \gamma
  s_{\mathrm{B}} + {1 \over \Delta'_{\mathrm{B}}}\fluxl\\ \meanf{N} &
  = & \fluxc{} + \gamma s_{\mathrm{N}} + {1 \over
  \Delta'_{\mathrm{N}}}\fluxl
\end{eqnarray}
This is a system of three linear equations with three unknowns
($\fluxc$, $\gamma$ and $\fluxl$) that can be resolved easily.

\subsubsection{Two filters and a constant continuum}
\label{sec:onenarrow}
In this case, we denote the filters by \broadband{} and \narrowband{}
and suppose that \narrowband{} is a narrow-band filter.

The fluxes inside the filters are:
\begin{equation}
  \meanf{\broadband}=\fluxc + {1 \over \Delta'_{\broadband}}\fluxl
  \qquad \qquad \meanf{\narrowband}=\fluxc + {1 \over
  \Delta'_{\narrowband}}\fluxl
  \label{eq:flujo2:b}
\end{equation}

We account for the different effective width of the filters
introducing the new parameter $\epsilon\equiv \Delta'_{\narrowband}
/\Delta'_{\broadband}$.  As {\narrowband} is a narrow-band filter,
$\epsilon \ll 1$.  The parameters of the emission line are as
follows:
\begin{eqnarray}
  \fluxl & = & \Delta'_{\narrowband}\left(\meanf{N}-\meanf{B}
  \right)\frac{1}{1-\epsilon} \label{eq:def:fl}\\ \fluxc & = &
  \meanf{B}\left(\frac{1-\epsilon\; \mathcal{Q}}{1-\epsilon}\right)\\
  \ew & =
  &\Delta'_{\narrowband}\left(\mathcal{Q}-1\right)\left(\frac{1}{1-\epsilon
  \; \mathcal{Q}}\right)
  \label{eq:def:ew}
\end{eqnarray}
being $\mathcal{Q} \equiv \meanf{\narrowband}/\meanf{\broadband}$.
The color of a source (with the color normalization of
Sect.~\ref{sec:description}) is directly $m_{\broadband} -
m_{\narrowband} = 2.5 \log \mathcal{Q}$.  

Equation \ref{eq:def:ew}
shows that the EW does not change linearly with $\mathcal{Q}$.  There
exists an upper limit to the color that a source can have due to the
presence of an emission line:
\begin{equation}
  \mathcal{Q}_{\mathrm{max}} = \lim_{\ew \to \infty} \mathcal{Q}(\ew) 
  = \lim_{\ew \to \infty} 
  \left(\frac{1+\frac{\ew}{\Delta'_{\narrowband}}}{1+
    \epsilon \frac{\ew}{\Delta'_{\narrowband}}} \right)
  = \frac{\Delta'_\broadband}{\Delta'_{\narrowband}}
  \label{eq:color:max}
\end{equation}

It is worth noting that observing an emission-line object with a
narrow-band filter increases the contrast between the emission line
and the continuum, as the object is brighter in the line than in the
continuum.  Let us assume that we have two objects with the same
magnitude in the broad-band, one with line of equivalent width $W$ (named L) 
and the other without emission (denoted C). 
The ratio of the fluxes of the two objects
measured in the narrow-band filter is the contrast parameter, defined
originally by \citet{1995AJ....110..963T}. Using Eq.~\ref{eq:flujo2},
we obtain:
\begin{equation}
  \frac{f(\mathrm{L})}{f(\mathrm{C})} =
  \frac{1+\frac{\ew}{\Delta'_{\narrowband}}}{1+
    \frac{\ew}{\Delta'_{\broadband}}}
  \label{eq:contraste}
\end{equation}
In the original contrast parameter of \citet{1995AJ....110..963T},
the width of the filter is characterized by its FWHM. For our 
Eq. \ref{eq:contraste} we, more accurately,
use the effective width as defined in Eq.~\ref{eq:def:ewidth}. 
The numeric expression of the contrast parameter is equal to the equation 
obtained by inverting Eq. \ref{eq:def:ew}.

There are two limit cases for these equations. When the width of the
broad-band is much greater than the narrow-band width, we can consider
that the line do not contribute to the broad-band filter.  The
parameter $\epsilon$ tends to zero and the equations can be written
as:

\begin{eqnarray}
  \fluxl & = & \Delta'_{\narrowbandL} \left(\meanf{\narrowbandL} -
  \meanf{\broadband} \right)\label{eq:def:fl:o}\\ \fluxc & = &
  \meanf{\broadband} \label{eq:def:fc:o}\\ \ew & =
  &\Delta'_{\narrowbandL}\left(\frac{\meanf{\narrowbandL}}{\meanf{\broadband}}
  -1\right)
  = \Delta'_{\narrowbandL}\left(\mathcal{Q}-1\right)
  \label{eq:def:ew:o}
\end{eqnarray}

The second limit case is when one filter do not cover the emission
line. Thus, the line do not contribute to the flux in that band and
the final equations are again Eqs.~\ref{eq:def:fl:o} -
\ref{eq:def:ew:o}.

\subsection{Multiple lines inside the filter}
In most cases of astrophysical interest, more than one line lies
inside the narrow-band filter.  For example, \HA{} plus \NII{}, and
\OIII{} plus \HB. The flux recovered from the narrow-band filter comes
from all the lines, although each line experiences a different filter
transmittance.
  
We write Eq.~\ref{eq:flujodelta} with three different lines (the case
of \HA{} and \OIII{}) and denote their fluxes and wavelength with
super index 1, 2 and 3:
\begin{equation}
  f(\lambda)=\fluxc{} + \fluxl^1\ \delta(\lambda-\lambda^1_z) +
  \fluxl^2\ \delta(\lambda-\lambda^2_z) + \fluxl^3\
  \delta(\lambda-\lambda^3_z)
  \label{eq:flujodelta3}
\end{equation}

and Eq.~\ref{eq:flujo2} is substituted by:
\begin{equation}
  \meanf{} = \fluxc{}+\frac{1}{\Delta'(\lambda^1_z)}\fluxl^1 +
  \frac{1}{\Delta'(\lambda^2_z)}\fluxl^2 +
  \frac{1}{\Delta'(\lambda^3_z)}\fluxl^3
  \label{eq:flujo3}
\end{equation}
The wavelengths of the three lines can be obtained from the redshift
(or from the wavelength) of one of them (for example, $\lambda^2_z$
and $\lambda^3_z$ as a function of $\lambda^1_z$). We suppose that
line 1 is the line we are interested on.

If the emission lines fluxes have known ratios, we can simplify the
equation:
\begin{equation}
  \meanf{}=\fluxc{}+\fluxl^1 \left( \frac{1}{\Delta'(\lambda^1_z)} +
  \frac{1}{\Delta'(\lambda^2_z)} \frac{\fluxl^2}{\fluxl^1} +
  \frac{1}{\Delta'(\lambda^3_z)}\frac{\fluxl^3}{\fluxl^1} \right)
  \label{eq:flujo4}
\end{equation}

The effective width of the filter is substituted by a combined width
($\Delta''$) that depends on the particular set of lines studied and
their ratios, that we denote $r_2=\frac{\fluxl^2}{\fluxl^1}$ and
$r_3=\frac{\fluxl^3}{\fluxl^1}$. The combined width is:
\begin{equation}
  \Delta''\left(\lambda_z^1, r_2, r_3 \right) \equiv
  \left(\frac{1}{\Delta'(\lambda^1_z)} +
  \frac{1}{\Delta'(\lambda^2_z)} r_2 + \frac{1}{\Delta'(\lambda^3_z)}
  r_3 \right)^{-1}
  \label{eq:combinedwidth}
\end{equation}
The equation developed previously to obtain fluxes and EW are valid to
obtain directly the flux of the main line (line with index 1) using
$\Delta''$ instead of $\Delta'$. The mean wavelength $\bar{\lambda_z}$ is altered
also. It is still computed using Eq. \ref{eq:mean2}, but substituting
$\Delta'$ by $\Delta''$.

The ratio of the combined width with the width computed in the single
line case represents the fraction of the total flux that comes from
the main line. We denote this ratio by $\phi$.
\begin{equation}
  \phi(\lambda_z^1, r_2, r_3) = \frac{\Delta''(\lambda_z^1, r_2,
  r_3)}{\Delta'(\lambda^1_z)}
  \label{eq:combinecover}
\end{equation}
The fraction $\phi$ depends on the shape of the filter and the
different line ratios of the lines. We can obtain a limit case of
$\phi$ when the filters are wide (this is the case for broad-band
filters).  With that assumption, $\Delta'(\lambda^1_z) \simeq
\Delta'(\lambda^2_z) \simeq \Delta'(\lambda^3_z)$ and
Eq.~\ref{eq:combinecover} can be written as:
\begin{equation}
  \phi_l(r_2, r_3) = (1 + r_2 + r_3)^{-1}
  \label{eq:combinecoverlimit}
\end{equation}   

In the following subsections we apply Eq.~\ref{eq:combinedwidth} -
\ref{eq:combinecoverlimit} for the particular cases of \HA{} and
\OIIIsin.

\subsubsection{\HA{} and \NIIsin}
\label{sec:hanii}
Both lines of the doublet {\NIIsin} are separated about 20 \AA{} from
{\HA} in the rest frame. In this case, $\lambda^1_z$ is the redshifted
wavelength of \HA{}, $\lambda^2_z$ is the redshift wavelength of
\NIIa{} and $\lambda^3_z$ is the redshifted wavelength of \NIIb{}.

The global ratio I(\NIIsin)/I(\HA) is related with metallicity
\citep{1998AJ....116.2805V,2002aj....123.2302m,
2002ApJS..142...35K,2004MNRAS.348L..59P} and AGN activity
\citep{1987ApJS...63..295V,2004ApJ...617...64S}, showing a great range
of variation between star-forming galaxies.

Historically, \HA{} surveys that could not separate \NIIsin{} from
\HA{} have used mean values to correct for the contribution of
\NIIsin{}. A commonly assumed mean value of the flux ratio
{I(\NIIb)/I(\HA)} is 0.32. This is the value reported by
\citet{1992ApJ...388..310K} and the UCM survey
\citep{1997ApJ...475..502G} \citep[see also]{2005A&A...429..851J}.

This particular mean value is widely used in studies of the SFR
based on \HA{}, where the lines of \NIIsin{} 
can not be resolved \citep[e.g.][]{1998ApJ...495..691T,1999ApJ...519L..47Y,
  2000PASJ...52...73I,2001A&A...379..798P,
  2003ApJ...586L.115F,2004ApJ...601..805U}.

The same line ratio is obtained from the Sloan Digital Sky Survey 
Data Release 4\footnote{http://www.mpa-garching.mpg.de/SDSS/DR4/}
\citep[SDSS DR4,][]{2006ApJS..162...38A} 
if we apply a lower cutoff in EW of 20\AA{} (the detection limit of the UCM
survey), typical of classical emission-line galaxy surveys.
\begin{figure}
  \includegraphics[bb=45 63 385 270]{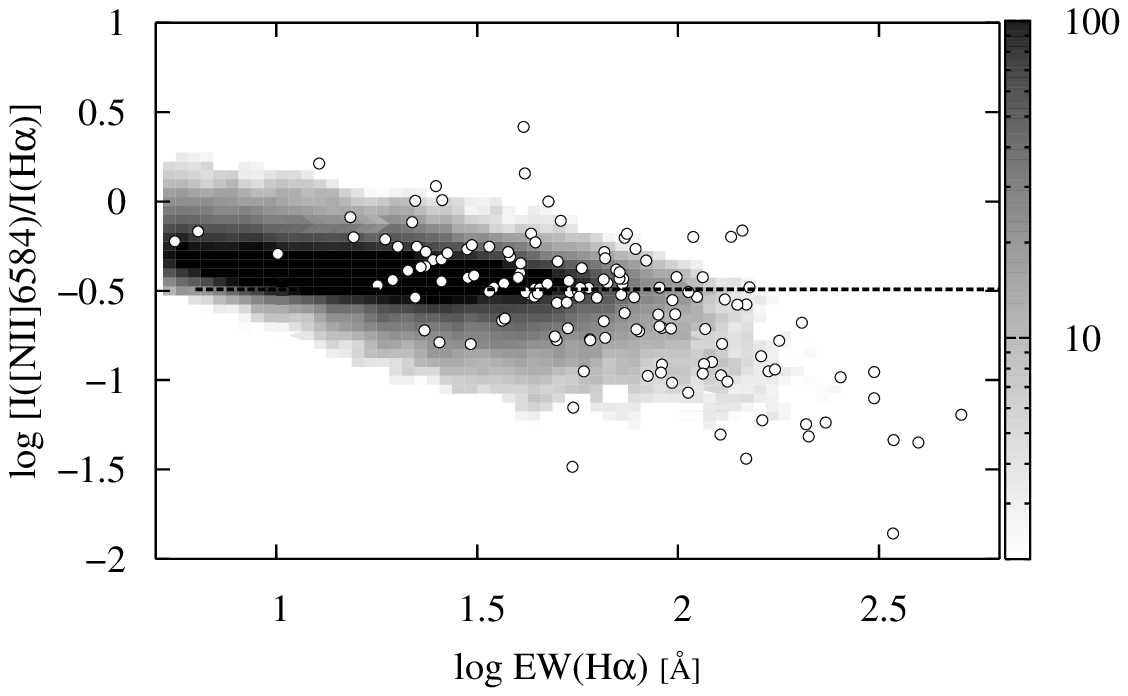}
  \caption{Ratio {I(\NIIb)/I(\HA)} represented as a function of the
    equivalent width of \HA{} for the galaxies of the UCM survey
    (white circles) and the Sloan Digital Sky Survey (the number of
    galaxies is represented in gray scale).  The mean value assumed in
    this work is represented as a straight line.  }
  \label{fig:niiha:intrinseco}
\end{figure}

In Fig.~\ref{fig:niiha:intrinseco} we show the ratio
{I(\NIIsin)/I(\HA)} represented as a function of the equivalent width
of {\HA} for the galaxies of the UCM survey and the SDSS.  The
equivalent width represented is observed, not rest frame.  As the
redshift of the main galaxy sample of SDSS is at most $z\sim$0.1, the
uncertainty produced by not correcting is about 4\%.  The effect in
the UCM is even less, as the galaxies in the sample are local.

The mean value assumed is represented as a straight line.  We can see
the wide range of variation of {I(\NIIsin)/I(\HA)}, from near 0.02 to
4.  There is a tendency of decreasing {I(\NIIsin)/I(\HA)} with
increasing EW, corresponding to a decrease of the metallicity.

The intensity of the nitrogen lines is related: I$(\NIIb)$/I$(\NIIa) =
3$ \citep{1989agna.book.....O}.

We can write Eq.~\ref{eq:combinedwidth} as a function of
{$\mbox{r=I(\NIIb)/I(\HA)}$} only.
In this case, the combined effective width of the filter is:
\begin{equation}
  \Delta''(\lambda^1_z, r) = \left(\frac{1}{\Delta'(\lambda^1_z)} +
  r\left(\frac{1}{3\,\Delta'(\lambda^2_z)} +
  \frac{1}{\Delta'(\lambda^3_z)}\right)\right)^{-1}
\end{equation}

We obtain the mean redshift of the sources using Eq. \ref{eq:mean2}
and inserting there the $\Delta''(\lambda_z, r)$. We can obtain also
the limit parameter $\phi_l = (1 + 4/3 r)^{-1}$.  With the mean value
assumed of $r=0.32$, $\phi_l= 0.71$. This means that, in the ideal
case of a square filter covering the three lines and the given line
ratios, 71\% of the flux recovered in Eq.~\ref{eq:def:fl} comes from
\HA{} and the rest comes from \NIIsin{}.

A narrow-band filter with the same $\phi$ value than the broad-band
filter is wide enough to contain simultaneously the two nitrogen lines
and \HA{}. A greater value of $\phi$ implies that the \NIIsin{} lines
only enter partially inside the narrow line filter. Using the standard
\NII{} correction with very narrow filters produce a systematic under
estimation of the line flux.
  
We can further refine the \NII{} correction. From
Fig~\ref{fig:niiha:intrinseco} we can obtain a relation between the
equivalent width \ew(\HA) and $r$=I(\NII)/I(\HA).  Using that relation
we can iterate through Eq.~\ref{eq:def:ew} and obtain simultaneously
$r$ and EW.  With this procedure, every object has a different value
of $r$, which implies different values of $\Delta''$ and $\phi$ for
every object.

\subsubsection{\HB{} and \OIIIsin}
\label{sec:hboiii}
In this case, three lines enter the filter, the doublet \OIIIsin{} and
\HB.  \HB{} is separated 146\AA{} from \OIIIb{} in the rest frame.  We
can write Eq.~\ref{eq:combinedwidth} with $\lambda^3_z$ as the
redshifted wavelength of \HB{}, $\lambda^2_z$ is the redshift
wavelength of \OIIIa{} and $\lambda^1_z$ is the redshifted wavelength
of \OIIIb{}.

\begin{figure}
  \includegraphics[bb=45 63 385 270]{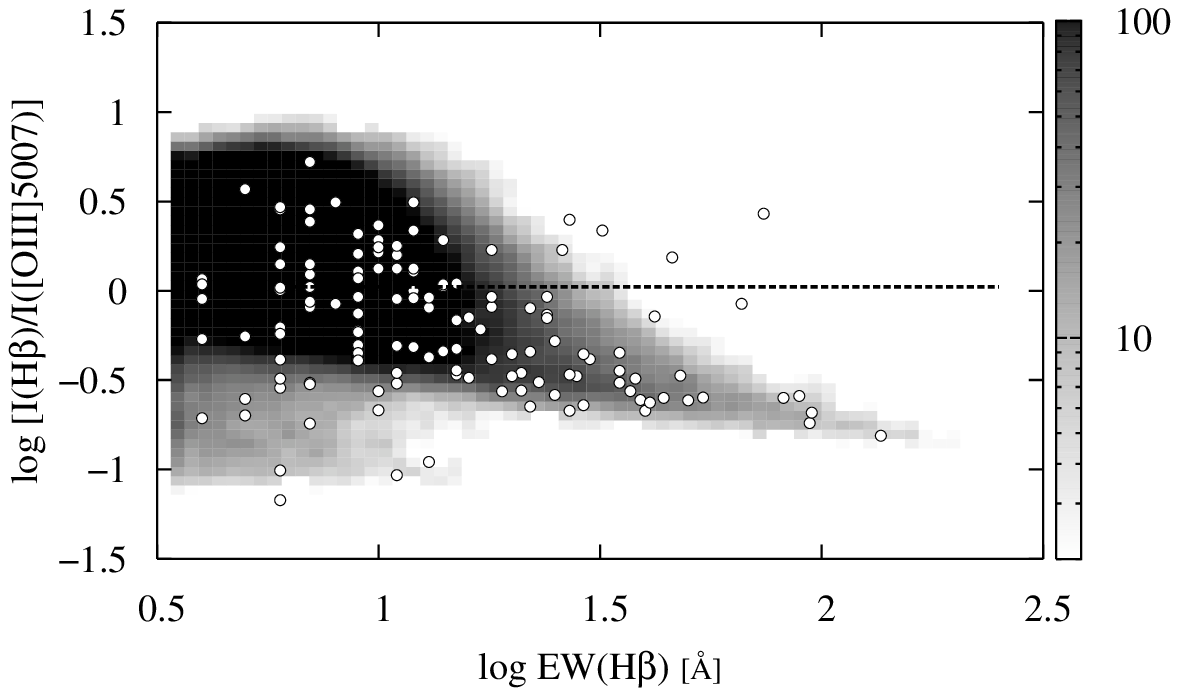}
  \caption{Ratio I({\HB})/I({\OIIIb}) represented as a function of the
    equivalent width of \HB{} for the galaxies of the UCM survey
    (white circles) and the Sloan Digital Sky Survey (the number of
    galaxies is represented in gray scale).  The mean value assumed
    in this work is represented as a straight line.}
  \label{fig:oiiihb:intrinseco}
\end{figure}

In Fig~\ref{fig:oiiihb:intrinseco} we show the ratio \OIIIb/\HB{}
again for galaxies of the SDSS DR4 and for galaxies of the UCM
sample. We have only represented galaxies with the lines \OIII, \HB{}
and a EW(\HA) $>$ 20\AA, in order to mimic the limit values of
classical ELG surveys.  The mean value obtained for I(\HB)/I(\OIIIb)
in the SDSS is $r=1.05$ (the result for the UCM is compatible,
$r=1.1$).
  
We write Eq.~\ref{eq:combinedwidth} for this particular case as a function 
of $r = \mathrm{I}(\HB)/\mathrm{I}(\OIIIb)$ and 
knowing that I$(\OIIIb)$/I$(\OIIIa) = 3$ \citep{1989agna.book.....O}:
\begin{equation}
  \Delta''(\lambda^1_z, r) = \left(\frac{1}{\Delta'(\lambda^1_z)} +
  \frac{1}{3\Delta'(\lambda^2_z)} +
  r\,\frac{1}{\Delta'(\lambda^3_z)}\right)^{-1}
\end{equation}

The limit parameter is in this case $\phi_l = (4/3 + r)^{-1}$. With
the mean value assumed for $r$, $\phi_l$ = 0.42, meaning that only
42\% of the flux would come from \OIIIb{}.

In this case, $\phi$ is very close to unity for the narrow-band
filters. This means that \OIIIb{} will enter alone in the filter. In
the case of broad-band, the ratio is close to the limit value, meaning
that the three lines enter the broad-band filter. As a summary, we can
consider that \OIIIb{} enters alone in the narrow-band filter, but we
have to include the three lines when considering the broad-band filter.

\subsection{Line luminosity of the objects}
\label{sec:lum}
The line luminosity of each object can be calculated from the measured
line flux and the redshift of the object (obtained from
$\bar\lambda_z$). The line flux corrected from obscuration by dust
($\fluxl^0$) can be obtained from the measured line flux ($\fluxl$):
\begin{equation}
  \log \fluxl^0 =\log \fluxl + 0.4 (A_{\lambda_r} + a_{\lambda_z})
\end{equation}
being $A_{\lambda_r}$ the internal obscuration at the rest frame
wavelength $\lambda_r$ and $a_{\lambda_z}$ the galactic obscuration.
The line luminosity is then:
\begin{equation}
  L = 4 \pi \; d^2_{\mathrm{L}}(\bar{z}) \;\fluxl^0
\end{equation}
being $d_{\mathrm{L}}(z)$ the luminosity distance.

The common obscuration corrections applied in narrow-band searches are
discussed in the following subsection.

\subsubsection{The obscuration correction}
\label{sec:obscuration}

Obscuration by dust affects seriously any measurement of galaxy
luminosities at the optical and UV wavelengths. Correcting for this
effect is a difficult task as measurements of parameters sensitive to
obscuration are not always easy to obtain.

In the case of obscuration corrections for emission lines, it is
common to assume an average value of 1$^m$ in {\HA} and the extinction
curve of \citet{1989ApJ...345..245C} to apply the value to other
emission lines.  This value of $A_{H\alpha}$ has been found as the
average of the obscuration of several samples of local {\HA} emitting
galaxies
\citep[e.g.][]{1983ApJ...272...54K,1995ApJ...455L...1G,2003ApJ...599..971H}.
This procedure has the advantage that is very simple to apply to a
luminosity function or luminosity density, since it is a scaling
factor.

It is worth to note here that the average obscuration applied in UV
studies is about $A_V=1^m$
\citep[e.g.]{1996MNRAS.281..847T,2000MNRAS.312..442S}.
In the GALEX far-ultraviolet (FUV), \citet{2005ApJ...619L..51B}
found a mean obscuration of $A_{FUV}=1.1^m$ for galaxies
selected in the GALEX near-ultraviolet band. This value is roughly
$A_V \simeq 0.4^m$.
Galaxies in the GALEX UV Imaging Atlas of Nearby Galaxies 
\citep{2004AAS...205.4201G,2006astro.ph..6440G}
have a mean obscuration, considering only spirals and irregulars,  
of $A_{FUV}=1.8^m$, corresponding to $A_V \simeq 0.7^m$
(Gil de Paz, private communication)
as this sample can be considered as selected in the optical.

The average obscuration of $A_{H\alpha}=1^m$ corresponds to a value of
about $A_V\simeq 1.22^m$ as noted by \citet{2004ApJ...615..209H}.
This disagreement in the obscuration is a selection effect of the
different (FUV, UV and \HA{}) samples.  Galaxies selected at redder
wavelengths can have greater internal obscurations than those
selected at the UV, where heavily absorbed systems will not be
selected.

The value of $A$ at the wavelength of the emission line
is obtained applying a given extinction curve.  We are assuming that
the continuum and the emission line are affected by the same
obscuration, so the EW does not need obscuration correction. This
last assumption is not true if we use the obscuration recipe of
\citet{1996ApJ...458..132C}, that is applied for strong starbursts.
However, normal galaxies and moderate starbursts can be assumed
as showing no obscuration correction in EW.

\section{Magnitude-color selection criterion}
\label{sec:dispq}
In this Section, we study the underlying statistics of the
magnitude-color diagram used to select ELGs in the most simple
scenario, a broad-band filter and a narrow-band filter. A sample
selected in broad-narrow color is directly selected in EW
(Eq.~\ref{eq:def:ew}).  Thus, ideally, a sample can be created
selecting the objects with color in excess of a minimum color
$(m_{\broadband}-m_{\narrowband})_{\mathrm{min}}$ that translates into
a minimum equivalent width EW$_{\mathrm{min}}$.
    
The dispersion in the magnitude-color diagram is dominated by the
uncertainties in the magnitudes in the faint region and by the
different colors of the objects in the bright region. Furthermore, the
mean value of the color, $\mu[m_{\broadband}-m_{\narrowband}]$,
depends on $m_{\narrowband}$, because different types of objects can
dominate the color at different narrow-band magnitude ranges.  The
method of selection has to account for the different sources of
dispersion of the color.  A solution is to compute the standard
deviation of the color dispersion and consider candidates, with a
level of signification $n_{\sigma}$, the objects with color:
\begin{equation}
  m_{\broadband}-m_{\narrowband} >
  \max\lbrace (m_{\broadband}-m_{\narrowband})_{\mathrm{min}},
  \mu[m_{\broadband}-m_{\narrowband}] + n_{\sigma}
  \sigma[m_{\broadband}-m_{\narrowband}]\rbrace
\end{equation}

The standard deviation of the color distribution, as a function of
the narrow-band magnitude, can be calculated from the distribution
of objects. In the following subsection, we infer an analytic
expression that can be useful when there are not enough points in
the data to calculate directly the dispersion.

\subsection{Standard deviation of the color distribution}
Let us assume a population of objects with the same SED, given by
$f(\lambda)$, so all the objects have the same color.  

In general, the source flux in analog to digital units (ADU) collected in a CCD
detector from an object are proportional to the integrated flux of the
source through the pass band and the exposure time $t$:
\begin{equation}
  N \propto t \int f(\lambda)\,T(\lambda)\,\lambda\,\ud \lambda
  \label{eq:counts1}
\end{equation}

Equation~\ref{eq:counts1} can be written in terms of \meanf,
$\lambda_0$ and $\Delta$ as follows\footnote{
Equation~\ref{eq:counts2} can be also written in terms of
$\lambda_{\mathrm{eff}}$ \citep[Eq. 24]{2003A&A...401..781F} as
follows:
  \begin{displaymath}
    N \propto t\, \tilde{f}\, T_{\mathrm{peak}}\,\Delta \, \lambda_{\mathrm{eff}}
  \end{displaymath}
  being
  \begin{math}
    \tilde{f} = \frac{\int f(\lambda)\,\mathcal{T}(\lambda)\, \ud
      \lambda}{\int \mathcal{T}(\lambda)\,\ud \lambda}
  \end{math}
  and
  \begin{math}
    \lambda_{\mathrm{eff}} = \frac{\int \lambda \,
      f(\lambda)\,\mathcal{T}(\lambda)\, \ud \lambda}{\int
      f(\lambda)\, \mathcal{T}(\lambda)\,\ud \lambda}
  \end{math}
  \\ We prefer using $\lambda_0$ to avoid the flux dependence of
  $\lambda_{\mathrm{eff}}$, that would complicate the equations of
  Section~\ref{sec:el}.  }:
\begin{equation}
  N \propto t\, \meanf \,\, T_{\mathrm{peak}}\,\Delta \, \lambda_0
  \label{eq:counts2}
\end{equation}
where $T_{\mathrm{peak}}$ is the maximum value of the transmittance.

The ratio $\mathcal{Q}=\meanf{\narrowband}/\meanf{\broadband}$ is proportional to the ratio
of ADUs in the same bands $\mathcal{R}$:
\begin{equation}
  \mathcal{R} \equiv \frac{N_{\narrowband}}{N_{\broadband}} =
  \mathcal{Q} \,
  \left(\frac{\lambda^{\narrowband}_0}{\lambda^{\broadband}_0}\right)
  \left(\frac{\Delta_{\narrowband}}{\Delta_{\broadband}}\right)
  \left(\frac{T^{\narrowband}_\mathrm{peak}}{T^{\broadband}_\mathrm{peak}}\right)
  \left(\frac{t_{\narrowband}} {t_{\broadband}}\right) = b\,
  \mathcal{Q} \left(\frac{t_{\narrowband}}{t_{\broadband}}\right)
       \label{eq:def:ratio}
\end{equation}
In general, $\mathcal{Q}$ is a function with a strong dependency on
the wavelength because of night sky emission lines. In the cases
covered in the present work, the filters are selected to exclude
strong emission features in the sky spectrum but that is not always
possible, such as when using narrow band IR filters designed to cover
emission lines of astrophysical interest at zero redshift like
Br$\gamma$ or molecular hydrogen.

We include the parameters depending of the transmittance of the filters in:
\begin{equation}
  b \equiv
  \left(\frac{\lambda^{\narrowband}_0}{\lambda^{\broadband}_0}\right)
  \left(\frac{T^{\narrowband}_\mathrm{peak}}{T^{\broadband}_\mathrm{peak}}\right)
  \left(\frac{\Delta_{\narrowband}}{\Delta_{\broadband}}\right)
\end{equation}

Eq.~\ref{eq:def:ratio} applies independently of the flux source.  We
denote the quantities referred to the sky background with a index $S$
and the quantities referred to a particular object by $O$. In
particular:
\begin{equation}
  \mathcal{R_\mathrm{S}} =
  \frac{N^\mathrm{S}_{\narrowband}}{N^\mathrm{S}_{\broadband}} = b\,
  \mathcal{Q_\mathrm{S}}
  \left(\frac{t_{\narrowband}}{t_{\broadband}}\right)
  \label{eq:def:ratio:sky}
\end{equation}

The standard deviation of $\mathcal{Q_\mathrm{O}}$ can be obtained from the
standard deviation of the ratio of the source flux in ADUs:
${\mathcal{R}_\mathrm{O}}$  as:
\begin{equation}
  \left(\frac{\sigma[\mathcal{Q_\mathrm{O}}]}{\mathcal{Q_\mathrm{O}}}\right)^2=
  \left(\frac{\sigma[\mathcal{R}_\mathrm{O}]}{\mathcal{R}_\mathrm{O}}\right)^2=
  \left(\frac{\sigma[N^\mathrm{O}_{\broadband}]}{N^\mathrm{O}_{\broadband}}\right)^2
  +
  \left(\frac{\sigma[N^\mathrm{O}_{\narrowband}]}{N^\mathrm{O}_{\narrowband}}\right)^2
    \label{eq:ratios}
\end{equation}
We do not include a covariance term. The covariance between the
sources flux in ADUs should be negligible because we are measuring
different parts of the spectrum with very different filters.

The terms in the right-hand side of the equation are the inverses of
the signal-noise ratio (SNR) We adopt here the signal to noise ratio
as computed by \citet{1996AJ....112.1302H} with a slightly changed
notation:
\begin{equation}
  \mathrm{SNR} \equiv \frac{N_\mathrm{O}}{\sigma
      [N_\mathrm{O}]}=\frac{N_\mathrm{O}}{\sqrt{G(N_\mathrm{O}+ A
      n_\mathrm{S}) + A G^2 R^2}}
  \label{eq:def:snr}
\end{equation}
$N_\mathrm{O}$ are the ADUs of the object in a given band,
$n_\mathrm{S}$ the best estimation of the sky background per pixel,
$A$ the area in pixels of the aperture used to measure photometry, $R$
is the readout noise in ADUs and $G$ the inverse of the gain $g$.  The
term involving the readout noise can be neglected when the exposures
are long enough to be background limited. so the dominating
contribution to the noise are the Poisson terms. Note that the
background limit can be difficult to reach if the narrow-band filter
is particularly narrow, as the airglow windows considered in the
present work are regions of low sky background.

Equation~\ref{eq:def:snr} depends on the ADUs of the object and on the
sky background measured in both bands.  We use Eq.~\ref{eq:def:ratio}
to translate the flux in the broad band into flux in the narrow band,
assuming a given color for the object ($\mathcal{Q_\mathrm{O}}$) and
for the sky background ($\mathcal{Q}_\mathrm{S}$).  Finally, we put
the equation as a function of the ADUs per unit time and pixel of the
sky ($\dot{n}_{\narrowband}^\mathrm{S}$) and of the ADUs per unit time
of the object ($\dot{N}^\mathrm{O}_{\narrowband}$), so we can easily
operate in magnitudes:
\begin{eqnarray}
  \left(\frac{\sigma[\mathcal{Q}_\mathrm{O}]}{\mathcal{Q}_\mathrm{O}}\right)^2
  & = &
  G\left(\frac{1}{t_{\narrowband}}+b\,\mathcal{Q}_\mathrm{O}\,
  \frac{1}{t_{\broadband}}\right)
  \frac{1}{\dot{N}^\mathrm{O}_{\narrowband}}+ \nonumber\\ &+&G\,
  q_\narrowband
  \left(\frac{1}{t_{\narrowband}}+b\,\frac{\mathcal{Q}_\mathrm{O}^2}
  {\mathcal{Q}_\mathrm{S}}\frac{1}{t_{\broadband}}\right)
  \frac{1}{\dot{N}^\mathrm{O}_{\narrowband}} + \nonumber\\ &+& A G^2
  R^2 \left(\frac{1}{t_{\narrowband}^2}+b^2 \,\mathcal{Q}_\mathrm{O}^2
  \frac{1}{t_{\broadband}^2}\right)
  \left(\frac{1}{\dot{N}^\mathrm{O}_{\narrowband}}\right)^2
       \label{eq:curve}
\end{eqnarray}

We introduce a new parameter $q$ that indicates if we are in the noise
regime dominated by the background or in the regime dominated by the
Poisson noise produced exclusively by the source flux.  The parameter
$q$ is the ratio between the flux in ADUs (or ADUs per unit time)
coming from the sky background and that coming from the object, inside
the given aperture $A$ and measured with a certain filter:
\begin{equation}
  q \equiv A \frac{n_{\mathrm{S}}}{N_\mathrm{O}} = A
  \frac{\dot{n}_{\mathrm{S}}}{\dot{N}_\mathrm{O}} = \frac{A}{p^2}\,
  10^{-0.4 (\mu - m)}
\end{equation}
where $\mu$ is the surface brightness of the sky, $m$ the magnitude of
the object and $p$ the plate scale.  $q$ tends to zero when the
background level is low compared with the brightness of the object and
to infinity when the background dominates over the object.

The previous equation is written in terms of $q_{\narrowband}$, but
can be rewritten easily in terms of $q_{\broadband}$, as
$q_{\narrowband} = q_{\broadband} \, \mathcal{Q}_\mathrm{S} /
\mathcal{Q}_\mathrm{O}$.

Equation \ref{eq:curve} shows $\sigma[\mathcal{Q}_\mathrm{O}]$ as a
function of $\dot{N}^{\mathrm{O}}_{\narrowband}$, and through it, as a
function of the narrow-band magnitude.  Of the three terms of the
equation, the last corresponds to the readout noise and is negligible
except for very short exposure. The second term comes from the contribution
of the sky background and dominates for faint objects. The first term
is produced by the ADUs of the object and only is important (when
compared with the sky contribution) for bright objects.
  
The presence of a logarithm in the transformation between flux and
magnitude produces an asymmetric distribution of points around the
mean color, mainly in the region of faint objects. We obtain this
behavior by defining two curves
\begin{eqnarray}
  (m_{\broadband} - m_{\narrowband})_{-n_{\sigma}} & 
  = & 2.5\log\left(\mathcal{Q} - n_\sigma\sigma[\mathcal{Q}]\right)\\
  (m_{\broadband} - m_{\narrowband})_{+n_{\sigma}} & 
  = & 2.5\log\left(\mathcal{Q} + n_\sigma\sigma[\mathcal{Q}]\right)
\end{eqnarray}
with $n_{\sigma}$ the level over the mean.

\begin{figure}
  \plotone{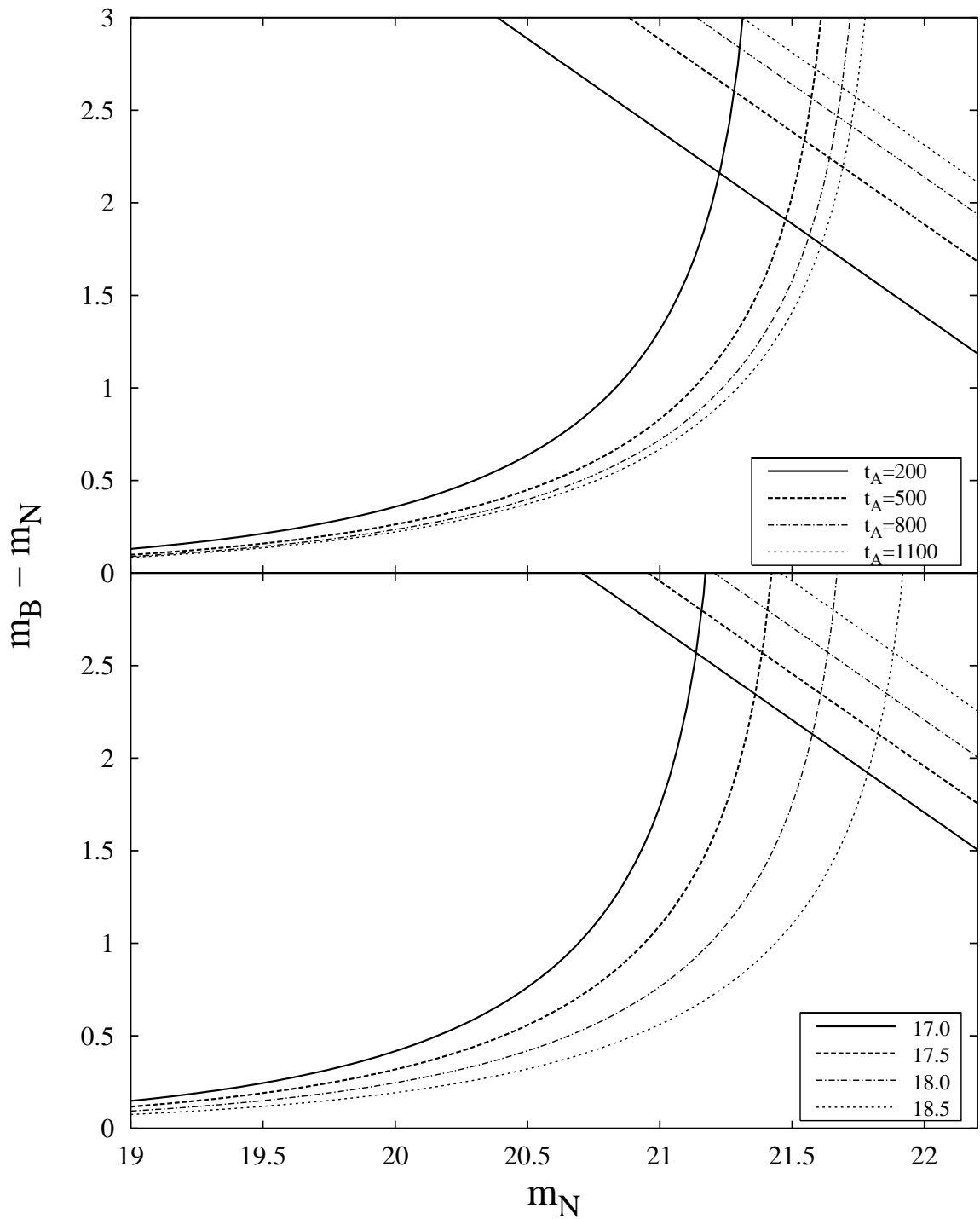}
  \caption{Selection curves with several different exposure times
    (top) and sky background (bottom).  The broad-band exposure times
    are: 200, 500, 800 and 1100 s, whereas the exposure time in the
    narrow-band filter is 800 s.  The sky background broad-band
    magnitudes are 17$^m$, 17.5$^m$, 18$^m$ and 18.5$^m$.  We
    represent also the 5-$\sigma$ broad-band limit magnitude (lines
    crossing on right top) corresponding to each exposure time and sky
    background magnitude.}
  \label{fig:seleccion2}
\end{figure}

As can be seen from Eq.~\ref{eq:curve}, for a given narrow-band
magnitude, $\sigma[\mathcal{Q}_\mathrm{O}]$ depends on the properties
of the filters encoded in $b$, on the exposure times in the two bands
($t_{\narrowband}$ and $t_{\broadband}$), on the color of the sky
background ($\mathcal{Q}_{\mathrm{S}}$) and on the relative brightness
of the sky when compared with the brightness of the objects ($q$).  In
Fig.~\ref{fig:seleccion2} we show how the selection curve changes
(with a fixed value of $n_\sigma$=3) with the exposure time and sky
background in the two bands for the sample filter set CAFOS8200 (the
different filter properties are detailed in Sect. \ref{sec:examples})
The increase of exposure time or the reduction of the
sky background (effectively increasing the sky-background magnitude)
reduces the slope of the selection curve.

\subsection{Isolines in the magnitude-color space}
\label{sec:sensibilidad}

Isolines represent curves joining points with an equal given
property. Lines with equal EW correspond with lines of equal color
in the magnitude-color diagram, but lines with equal continuum or
line flux are not directly related with neither the color nor the
magnitude of any of the bands.

\subsubsection{Isolines of line flux}

We can compute the isolines in magnitude-color space starting 
with Eq.~\ref{eq:flujo2}. If we assume a constant line flux $\fluxl$, the
magnitudes of the objects as a function of $\fluxl$ and EW only are:
\begin{eqnarray}
  m_{\broadband}&=& C-2.5\log\left[\fluxl \left(
    \frac{1}{\ew}+\frac{1} {\Delta'_{\broadband}}\right) \right]\\
    m_{\narrowband}&=& C-2.5\log \left[ \fluxl \left(
    \frac{1}{\ew}+\frac{1}{\Delta'_{\narrowband}}\right)\right]\\
    m_{\broadband} - m_{\narrowband} & = & 2.5\log
    \left(\frac{1+\frac{\ew}{\Delta'_{\narrowband}}}{1+
    \frac{\ew}{\Delta'_{\broadband}}} \right)\label{eq:iso:color}
\end{eqnarray}
The luminosity of the emission line is proportional to
the measured flux (see Sect.~\ref{sec:lum} for details), so the
isolines of line flux are also isolines of line luminosity.

We can write a parametric expression for the isolines of line flux.
We use $w=\ew/\Delta'_{\narrowband}$ as an adimensional
parameter. Note that the equations are also valid for absorption lines
(\fluxl $< 0$ and $\ew < 0$). To simplify the notation, we rename the
color $m_{\broadband} - m_{\narrowband}$ as the independent variable
$y$ and the narrow-band magnitude as $x$.

The parametric equations of the line flux isolines are:
\begin{eqnarray}
  x &=& x_{\infty} -
  2.5\log\left|1+\frac{1}{w}\right|\label{eq:isofl:param1} \\ y &=&
  2.5\log\left(\frac{1+w}{1+\epsilon w}\right)
  \label{eq:isofl:param}
\end{eqnarray}
where $x_{\infty}$ is the narrow-band magnitude of an object dominated
by the emission line:
\mbox{$x_{\infty}=C-2.5\log\left|\frac{\fluxl}{\Delta'_{\narrowband}}\right|$}. Equations
\ref{eq:isofl:param1} and \ref{eq:isofl:param} are valid in the ranges
of emission lines, corresponding to $w \in (0, \infty)$, and in the
domain of absorptions lines, where the adimensional parameter $w \in
(-1, 0)$.

The asymptotic behavior of the curve is as follows.  For an object
without line $w \to 0$; the color $y \to 0$.  When $w$ increases
asymptotically the object is dominated by the emission line.  Then , $x
\to x_{\infty}$ (the magnitude of an object dominated by the emission
line) and the color $y \to -2.5\log \epsilon$.  When $w \to -1$, the
object is dominated by the absorption line; the absorption line grows
stronger and the narrow-band magnitude tends to positive infinity; the
object is progressively dimmer as the absorption line increases its
equivalent width. The broad-narrow color tends to negative infinity.


\subsubsection{Isolines of continuum flux}

The isolines of continuum flux can also be obtained:
\begin{eqnarray}
  m_{\broadband}&=&C - 2.5\log \left[ \fluxc \left(1+
    \frac{\ew}{\Delta'_{\broadband}}\right) \right]\\
  m_{\narrowband}&=&C - 2.5\log \left[ \fluxc \left(1+
    \frac{\ew}{\Delta'_{\narrowband}}\right) \right]
\end{eqnarray}

The equation of the color $m_{\broadband}-m_{\narrowband}$ is equal to
Eq. \ref{eq:iso:color}, because the color depends only on the
equivalent width.  A similar approach to obtain the equation of the
isolines of continuum flux can be used. The adimensional parameter is
again $w= \ew /\Delta'_{\narrowband}$ and the parametric equations
are:
\begin{eqnarray}
  x &=& x_0 - 2.5\log\left(1+w\right)\\
  y &=& 2.5\log\left(\frac{1+w}{1+\epsilon w}\right)
  \label{eq:isofc:param}
\end{eqnarray}
being $x_0=C-2.5\log\left(\fluxc\right)$ the magnitude on an object
without line and $w \in (-1, \infty)$.  The asymptotic limits
correspond to a object without line ($w \to 0$, $x \to x_0$ and $y \to
0$), an object dominated by the emission line ($w \to \infty$, $x \to
-\infty$ and $y \to -2.5\log \epsilon$) and an object dominated by an
absorption line ($w \to -1$, $x \to \infty$ and $y \to -\infty$)
  
\begin{figure}
  \plotone{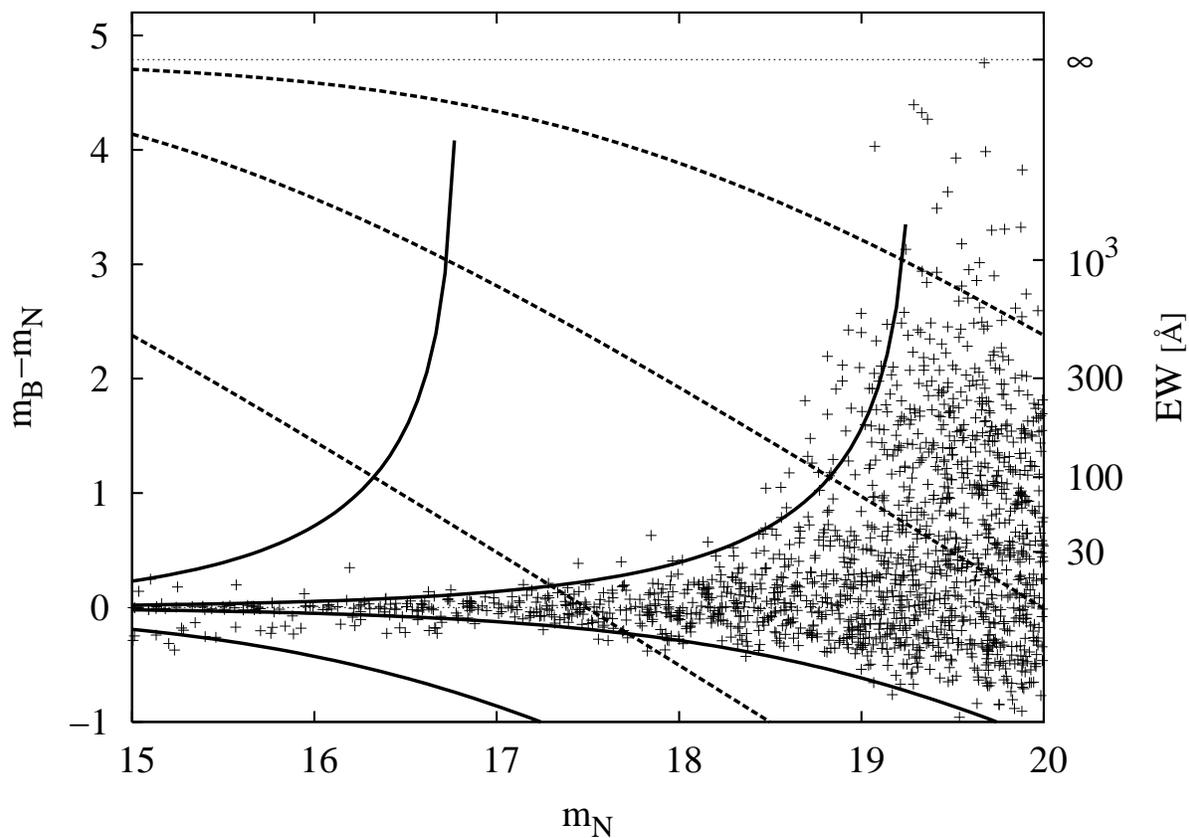}
  \caption{Isolines of line flux (solid curve) and continuum flux
    (dashed curve) for emission lines (broad-narrow color above zero)
    and absorption lines (bellow zero) in images obtained with the
    WFC9200 system (see Sect.~\ref{sec:examples} for details).  The
    flux lines correspond to 10$^{-14}$ and 10$^{-15}$ erg
    s$^{-1}$. The continuum lines are 10$^{-16}$, 10$^{-17}$ and
    10$^{-18}$ erg s$^{-1}$ \AA$^{-1}$.}
  \label{fig:isolineas}
\end{figure}

In Fig.~\ref{fig:isolineas}, we show a magnitude-color selection
diagram including example isolines for the WFC9200 case 
(see Sect.~\ref{sec:examples} for details).  The isolines
of continuum flux and line flux are represented by the dashed and
solid curves respectively. We can see clearly the asymptotic behavior
of both types of isolines. Although the limited numeric precision of
the plot do not show it, the curves of line flux continue up to the
maximum color.
  
\subsection{Redshift range of detectability and covered volume}
  
We have assigned an approximate redshift $\bar{z}$ to the source, but
we do not know which is the redshift range $z_1$ $z_2$ of the objects
that could be selected and included in the sample.  The answer to this
question is the key for the incompleteness correction. If we intend to
compute a luminosity function, we need to compute the comoving volume
per object.
    
We consider that the object is at redshift $\bar{z}$ and it has a line
flux $\fluxl$ and continuum flux $\fluxc$.  The flux recovered in a
given filter at redshift $z$ (whereby the emission line lies at
wavelength $\lambda_z$ ) is:
\begin{equation}
  f(z)=\left(\fluxc \left(\frac{1+\bar{z}}{1+z}\right) +
  \frac{\fluxl}{\Delta'(\lambda_z)}\right)\left(\frac{d_L(\bar{z})}{d_L(z)}\right)^2
  \label{eq:fluxz}
\end{equation}
with $d_L$ the luminosity distance. For $z=\bar{z}$ we recover
Eq. \ref{eq:flujo2}.  Substituting the appropriate filter widths
$\Delta'$, we can obtain the magnitude and color of the object as a
function of redshift. Only at a range of redshifts $z_1, z_2$ the
object fulfills the color criterion given in
Sect.~\ref{sec:dispq}.

\begin{figure}
  \plotone{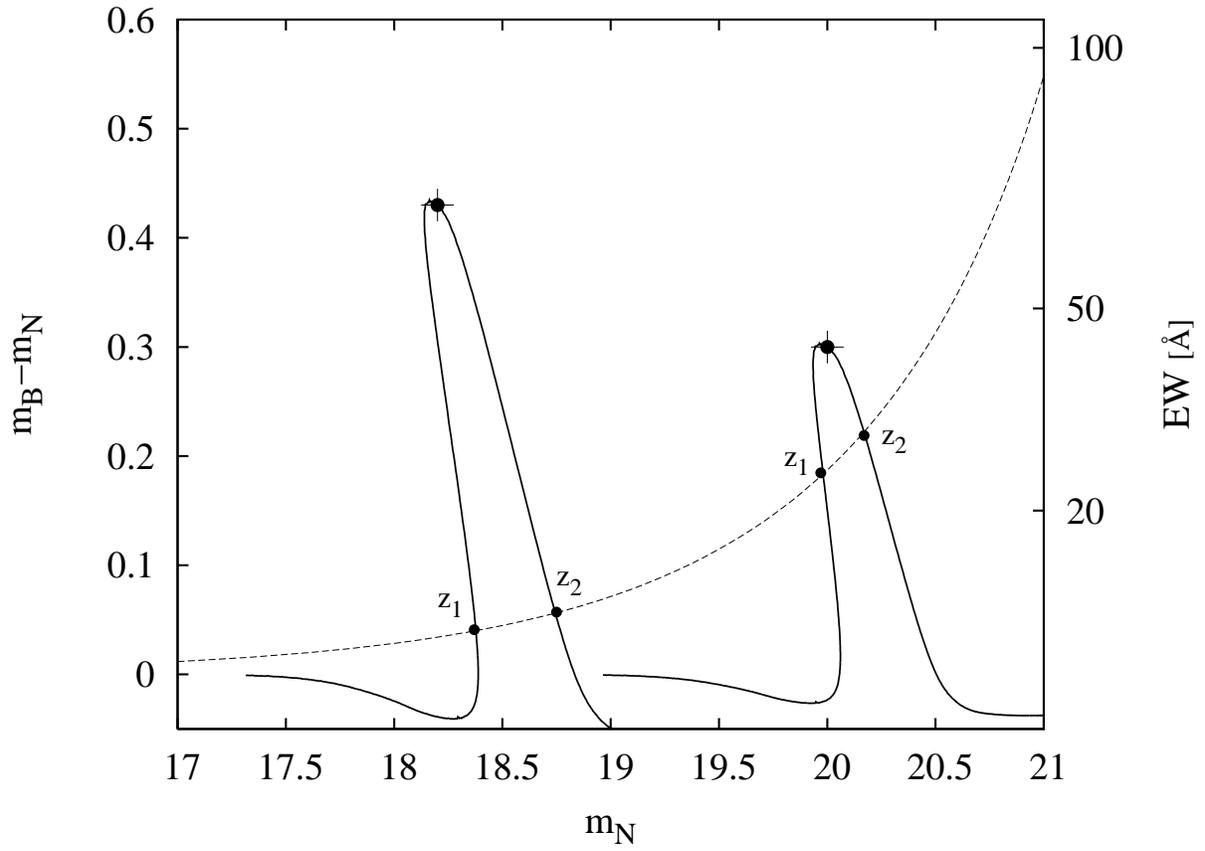}
  \caption{Redshift traces (solid lines) of two objects marked as
    crosses in a magnitude-color diagram for CAFOS8200 (see
    Sect.~\ref{sec:examples} for details). The emission line in this
    case is \HA{}. The selection limit is the dashed line. The
    intersection between the traces and the selection curve delimits
    the redshift range (from $z_1$ to $z_2$ in the plot) where the
    object could still be selected.}
  \label{fig:volume}
\end{figure}

In Fig.~\ref{fig:volume} we show the redshift traces given by
Eq.~\ref{eq:fluxz} for two objects in a magnitude-color diagram
corresponding to the filter set CAFOS8200 (see
Sect.~\ref{sec:examples} for details). The position of the objects in
the diagram at redshift $\bar{z}$ are represented as crosses on top of
the traces.  The intersection between the traces and the selection
curve (drawn with a dashed line) is labeled as $z_1$ and $z_2$. The
object would be selected only in the range $z_1 < z < z_2$.

The values of $z_1$ and $z_2$ depend through Eq.~\ref{eq:fluxz}, on
the transmittance of the filters. As the transmittance of the
broad-band filter is mainly constant in the wavelength range of
interest, the narrow-band filter dominates.  Thus, the redshift range
$z_1, z_2$ is small and dominated by the shape of the transmittance of
the narrow-band filter.  The redshift range depends also on the
selection curve.  Objects selected at lower signification level are
closer to the selection limit and the redshift range $z_1, z_2$ is
smaller.

The comoving volume where the object could be selected is computed
from the redshifts $z_1, z_2$.  This volume is one of the ingredients
needed to compute the luminosity function.  With Eq.~\ref{eq:fluxz}
and the selection curve, we can compute a comoving volume for each
object. It can be seen clearly that, with the same EW, faint objects
have smaller comoving volumes. These objects have, for a fixed EW, low
line flux (and low line luminosity). Thus, assuming a global comoving
volume for all the detected objects tends to underestimate the value
of the luminosity function for the faint objects and objects selected
near the selection curve.

\section{Example filters}
\label{sec:examples}
\subsection{Sets of filters and instruments}
  
In order to provide observational tests for our analysis, we select
three sets of filters in two different telescopes.  The first
instrument is CAFOS on the 2.2 m telescope at Calar Alto Observatory
(Almer\'{\i}a, Spain). The filters used were 816/16 (narrow band) and
850/150c (broad band, Johnson $I$). This is the instrumental setup
used in the \citet{2001A&A...379..798P} survey. The second instrument
is the Wide Field Camera \citep[WFC,][]{1996SPIE.2654..266I}, located
at the Isaac Newton Telescope in the \emph{Observatorio del Roque de
los Muchachos} (island of La Palma, Spain). Two different narrow-band
filters have been used, \#209 and \#208, together with the broad band
filter \#194 (RGO $I$).

\begin{figure}
  \plotone{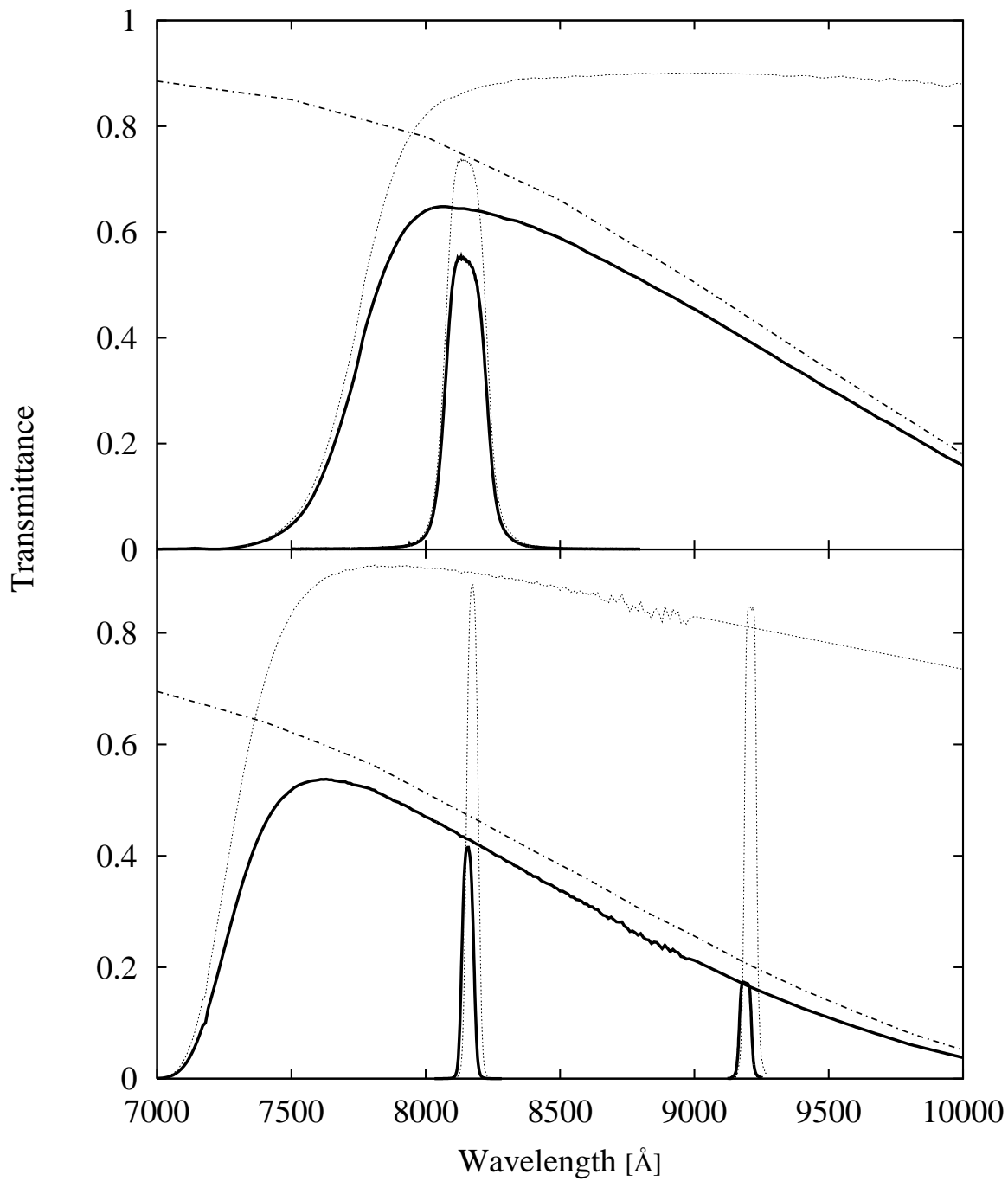}
  \caption{Transmittance curve for the different sets of
    filters. Dash-dotted lines represent the quantum efficiency,
    dotted lines represent the fiducial transmittance of the filters.
    Finally, the solid lines represent the real transmittance
    corrected from the different contributions detailed in the
    text. On top, the filters from CAFOS. Broad-band is 850/150c and
    narrow-band 816/16.  On bottom the filters from WFC. Broad-band is
    \#194 and narrow-band are \#209 around 8200{\AA} and \#208 around
    9200{\AA}.  }
  \label{fig:filters}
\end{figure}

The transmittance of the different filters is shown in
Fig. \ref{fig:filters}.  The fiducial transmittances are displayed
together with the QE of the detector and the total transmittance.  For
a working temperature of 0\degr C, the blue shift produced by the
temperature and the converging beam for CAFOS8200 is 10\AA, about 5\%
of the width of the filter (S. Pedraz, private communication).  The
blue shift is 17 \AA{} for the filter WFC8200 and 19 \AA{} for the
filter WFC9200 \citep{spectrum}. These two shifts are about $\sim$35\%
of the width of the filters. The converging beam can produce also an
effect in the absolute transmittances and widths, but it turned out to
be negligible for our examples.

Several quantities useful to describe filters are summarized in
Table~\ref{tab:properties}. Columns 1 and 2 are the name of the filter
set and the name of the filter (either broad- or narrow-band). In the
next three columns we show $\lambda_0$ (column 3), $\sqrt{\mu^2}$
(column 4) and $\Delta$ (column 5) computed directly from the
transmittance. Column 6 is the estimated mean value of $\lambda_z$
computed with Eq.~\ref{eq:mean2}, concordance cosmology and the \HA{}
line, not including the corrections for the \NIIsin{} lines (see
bellow).  With $\bar\lambda_z$ we compute the effective width of the
filter in column 7.  The difference between commonly used cosmologies
and different emission lines is less than 5\% in all cases.  Note that
the filter \#194, when used together with \#208 has an effective width
roughly three times its width. The wavelength where the emission line
is located has low transmittance. In this case, using $\Delta$ instead
of $\Delta'$ would underestimate the continuum flux.

\begin{deluxetable}{llccccc}
  \tablewidth{0pt} \tablecaption{Summary of properties of the filters
  used
    \label{tab:properties}
  } \tablehead{ \colhead{System} & \colhead{Filter} &
  \colhead{$\lambda_0$} & \colhead{$\sqrt{\mu^2}$} &
  \colhead{$\Delta$} & \colhead{$\bar\lambda_z$} &
  \colhead{$\Delta'(\bar{\lambda_z})$}\\ & & \colhead{(\AA)} &
  \colhead{(\AA)} & \colhead{(\AA)}&\colhead{(\AA)} &
  \colhead{(\AA)}\\ \colhead{(1)} & \colhead{(2)} & \colhead{(3)} &
  \colhead{(4)} & \colhead{(5)}& \colhead{(6)} & \colhead{(7)}\\ }
  \startdata CAFOS8200 & 816/16 (narrow) & 8139 & 80 & 173 & 8147 &
  176\\ & 850/150c (broad) & 8715 & 643 & 1665 & 8147 & 1783 \\
  WFC8200 & \#209 (narrow) & 8157 & 17 & 48 & 8157 & 48\\ & \#194
  (broad) & 8007 & 477 & 1333 & 8157 &1940\\ WFC9200 & \#208 (narrow)
  & 9190 & 17 & 49 & 9190 & 49\\ & \#194 (broad) & 8007 & 477 & 1333 &
  9190 & 4398\\ \enddata
\end{deluxetable}

In Table~\ref{tab:width:mline}, we summarize the mean values of the
combined width and $\phi$ parameter (Eq ~\ref{eq:combinecover} for
\HA{} and \OIIIsin-computed quantities with concordance cosmology.
Columns 1 and 2 have the same meaning in Table~\ref{tab:properties}.
Columns 3 to 5 correspond to \HA-computed quantities with
$r=${I(\NIIb)/I(\HA)}=0.32 (Sect. ~\ref{sec:hanii}) and columns 6 to 8
to \OIIIsin{} with $r=$I(\HB)/I(\OIIIb)=1.05 (Sect.\ref{sec:hboiii}).
Columns 3 and 6 are the mean wavelengths computed according to
Eq~\ref{eq:mean2}, using the value of $\Delta''$ . Columns 4 and 7 the
value of the combined width at the computed mean wavelength value and
columns 5 and 8 the parameter $\phi$, that denotes the the fraction of
the total flux that comes from the main line.

For \HA{}, the narrow-band filter in the CAFOS8200 filter set has the
same $\phi$ value than the broad-band filters. This means that is wide
enough to contain simultaneously the two nitrogen lines and \HA{}.
The narrow-band filters of WFC8200 and WFC9200 have a greater $\phi$,
meaning that the \NIIsin{} lines only enter partially inside the
narrow line filter.  In this two last cases, 82\% and 85\% of the flux
in the narrow-band filter come from \HA{}, instead of the nominal
71\%. Using the standard \NII{} correction with very narrow filters
such as the narrow-band filters of WFC, produce a systematic under
estimation of the line flux of about 15\%.

In the case of \OIIIsin{}, $\phi$ is very close to unity for the
narrow-band filters. This means that \OIIIb{} will enter alone in the
filter. In the case of broad-band, the ratio is close to the limit
value (0.42 with the assumed value of $r$), meaning that the three
lines enter the broad-band filter. As a summary, we can consider that
\OIIIb{} enters alone in the narrow-band filter, but we have to
include the three lines in the broad-band filter.

\begin{deluxetable}{cccccccc}
  \tablewidth{0pt} \tablecaption{Mean combined effective widths of the
  filters\label{tab:width:mline}} \tablehead{ \multicolumn{2}{c}{} &
  \multicolumn{3}{c}{\HA{} + {\NIIsin} with $r$=0.32} &
  \multicolumn{3}{c}{\HB{} + \OIIIsin{} with $r$=1.05}\\
  \colhead{System} & \colhead{Filter} & \colhead{$\bar\lambda_z$} &
  \colhead{$\Delta''(\bar{\lambda_z})$} & \colhead{$\phi$} &
  \colhead{$\bar\lambda_z$} & \colhead{$\Delta''(\bar{\lambda_z})$} &
  \colhead{$\phi$}\\ & & \colhead{(\AA)} & \colhead{(\AA)} & &
  \colhead{(\AA)}&\colhead{(\AA)}\\ \colhead{(1)} & \colhead{(2)} &
  \colhead{(3)} & \colhead{(4)} & \colhead{(5)} & \colhead{(6)} &
  \colhead{(7)} & \colhead{(8)} }

  \startdata 
  CAFOS8200 & 816/16 (narrow) & 8142 & 123.5 & 0.70 & 8154 & 140.1 & 0.80\\ 
  & 850/150c (broad) & 8142 & 1262 &  0.71 & 8154 & 763 & 0.43\\
  WFC8200 & \#209 (narrow) & 8153 & 39.3 & 0.82 & 8157 & 47.5 & 0.99\\
  & \#194 (broad) & 8153 & 1375 & 0.71 & 8157 & 779 & 0.40\\
  WFC9200 & \#208 (narrow) & 9185 & 43.4 & 0.89 & 9190 & 45.8 & 0.93\\ 
          & \#194 (broad)  & 9185 & 3109 & 0.71 & 9190 & 1583 & 0.36\\
  \enddata
\end{deluxetable}

\subsection{Colors of galaxies}
We use galaxy templates to calculate the color evolution of different
galaxies with increasing redshift. With the set of templates of
\citet{1996ApJ...467...38K}, we have obtained colors for the three
groups of sample filters. The templates have a resolution of 5\AA. We
compare three wide types of galaxies with similar colors in this
particular system: ellipticals and bulges, early-type spirals are
represented by a Sb and finally late-type spirals and starbursts are
represented by the starburst template SB1.  The Sa and Sb spectra
contain the \HA{} line in emission. The Sc and the starburst spectra
also contain the nebular lines \OII, \HB{} and \OIII.

\begin{figure}
  \includegraphics[width=0.7\hsize,bb=50 50 482 776]{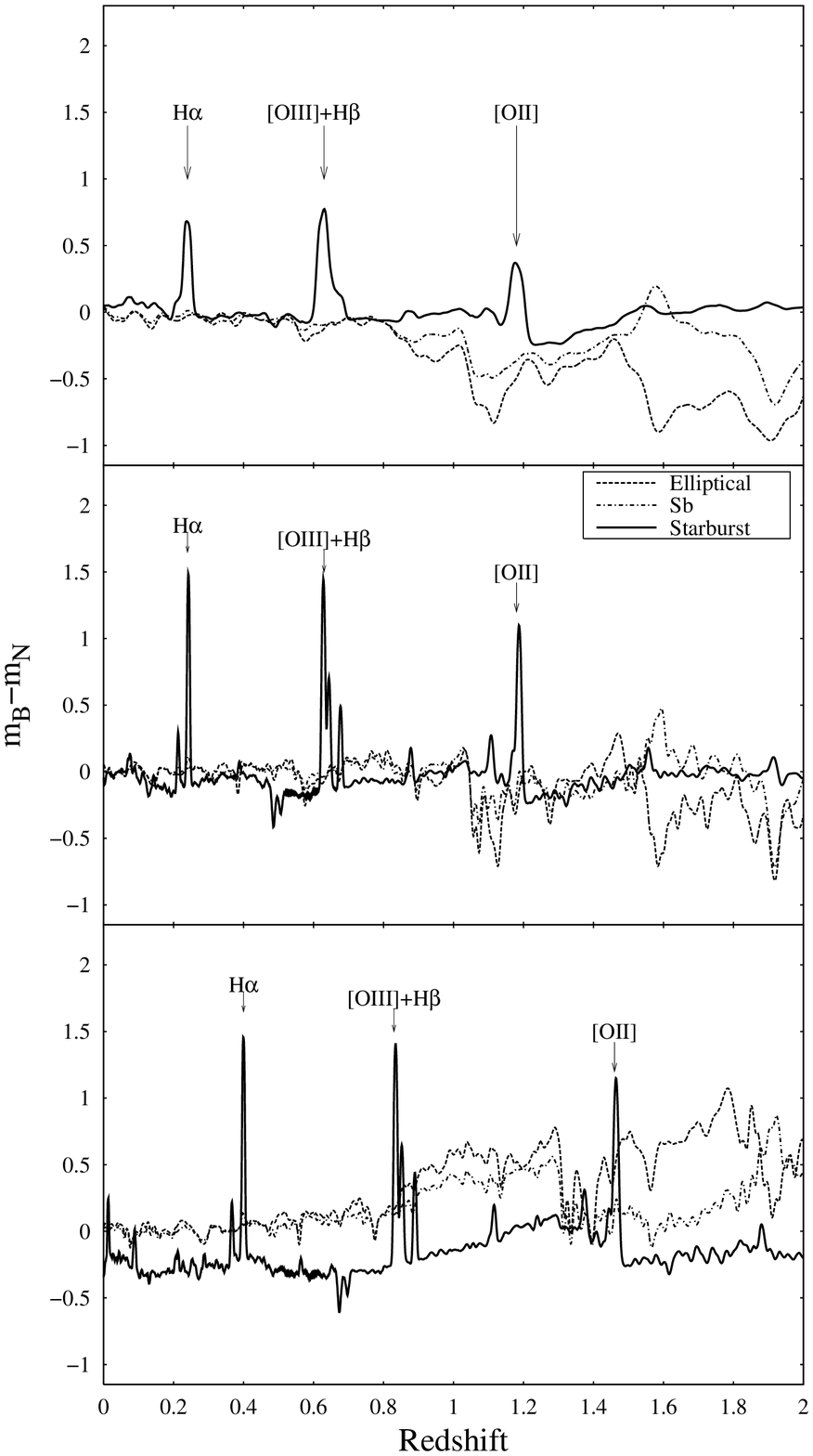}
  \caption{Colors of galaxy SEDs for the sample filter sets as a
    function of redshift. From top to bottom CAFOS8200, WFC8200 and
    WFC9200. Hubble types templates from \citet{1996ApJ...467...38K}
    are plotted. Ellipticals (and bulges) are represented by dashed
    lines, we use Sb template for early type spirals and S0
    (dash-dotted lines) and SB1 starbursts template for starburst and
    late spirals (Sc) (solid line). The position of the different
    emission lines is labeled.}
  \label{fig:color:gal}
\end{figure}

We show the color $m_{\broadband}-m_{\narrowband}$ of the different
templates as a function of redshift up to $z=2$ in
Fig.~\ref{fig:color:gal} for the three filter sets.  Only the effect
of the redshift is considered. We are not including galactic
evolution. As expected, galaxies with different emission lines show
strong increases on their colors when the emission lines enter the
region covered by the filter. In the filter set WFC9200, elliptical
templates exhibit color excess at redshifts around $z\sim1$ and higher.
This excess is not produced by emission lines. Instead, the very different
central wavelengths of the two filters involved and the slope of the continuum
near the 4000\AA{} break explains the broad-narrow color obtained.
  
\begin{figure}
\plotone{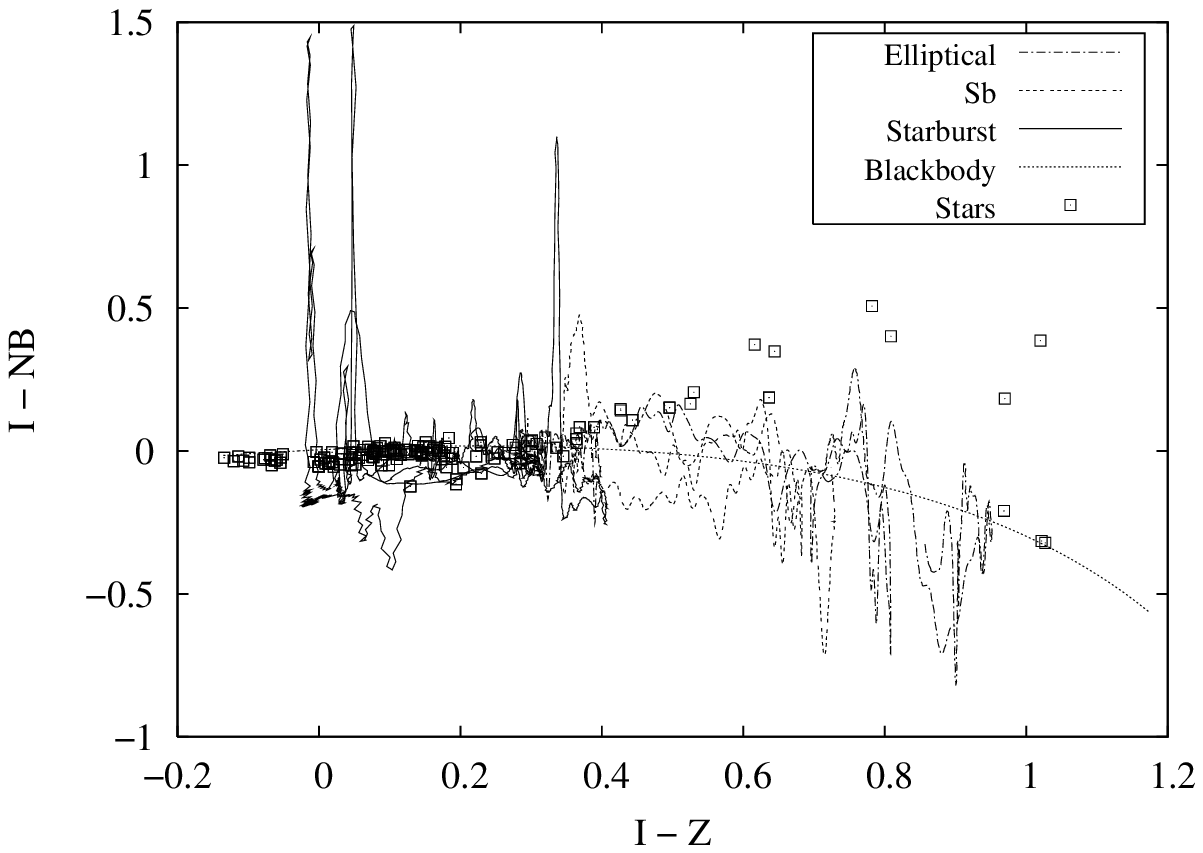}
\caption{$I$ - NB versus $I-Z$ color-color diagram for the WFC8200
  filter set. The key for the different types is represented in the
  box. Starbursts and late type stars exhibit color excess in $I$-NB
  but have very different colors in $I$-$Z$.  }
\label{fig:color-color}
\end{figure}

An additional color can help to distinguish different types of objects
with narrow-band color excess.  Figure~\ref{fig:color-color} shows a
\mbox{$I$ - NB} versus $I-Z$ color-color diagram for the WFC8200
filter set. We represents simultaneously stars, the black-body and
galaxy templates.  With a \mbox{$I$ - NB} versus NB diagram we can not
distinguish between the color excess produced by the emission line of
a starburst galaxy and by the molecular bands of a late-type star. But
these objects have different $I-Z$ color. Late type stars are very
red, $I-Z >$ 0.4 and starbursts are blue, with $I-Z < 0.4$. Thus, we
can classify objects using their $I-Z$ color.

\subsection{Colors of stars}
Using the library of
\citet{1998PASP..110..863P}\footnote{http://www.ifa.hawaii.edu/users/pickles/AJP/hilib.html}
we have estimated the colors of stars of different spectral types in
our sample sets.  The library contains spectral energy distributions
of stars with a large range in metal content. The spectral range
covers up to 10000~{\AA}.

\begin{figure}
  \plotone{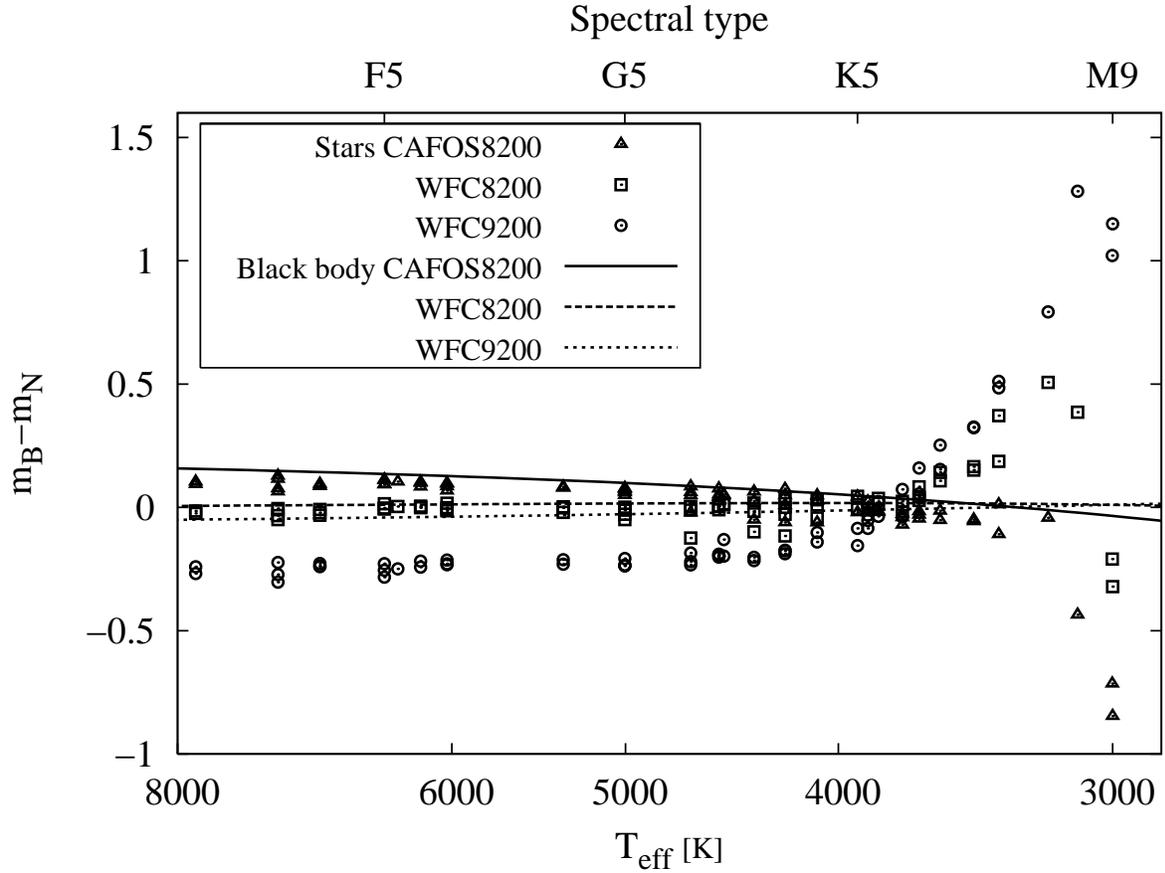}
  \caption{Colors of stars of late spectral types in our three sample
    filters. Stars of late spectral types are also represented with
    triangles, squares and circles for CAFOS8200, WFC8200 and
    WFC9200. The colors of a black body on the same filters have been
    added for comparison and are represented by solid, dashed and
    dotted lines respectively. The bottom axis represents the
    effective temperature of the star obtained from the spectral type
    of the model (top axis).}
  \label{fig:color:stars}
\end{figure}

In Fig.~\ref{fig:color:stars}, the color
$m_{\broadband}-m_{\narrowband}$ as a function of spectral type is
plotted. As noted before, stars follow the trend the colors of are
dominated by bands of TiO and VO.

Comparing the colors obtained with CAFOS8200 and WFC8200 we can see
that cold stars with molecular bands can be misclassified as ELGs 
in the case of WFC8200 and WFC9200.
    
\section{Optimal exposure times}
\label{sec:optimal}
A critical question that arises when planning observations with
narrow-band filters is how long we should expose in the narrow band
filter in order to obtain images that can be compared with the
corresponding broad band.  Exposure simulators available in
telescopes usually only cope with broad-band filters.

To be selected in a magnitude color diagram with a signification
$n_{\sigma}$, an object with an emission line has to show color excess
over the mean distribution of objects at a given magnitude
$m_\narrowband$.  The ratio of fluxes $\mathcal{Q}_\mathrm{l}$ of an
object with emission line has to fulfill:
\begin{equation}
  \mathcal{Q}_\mathrm{l} > \mu[\mathcal{Q}_\mathrm{O}] + n_{\sigma} \,
  \sigma[\mathcal{Q}_\mathrm{O}]
\end{equation}
being $\mathcal{Q}_\mathrm{O}$ the ratio of fluxes of the objects
without emission line and the same narrow-band magnitude that the
object with the emission line.
  
We can approximate $\mu[\mathcal{Q}_\mathrm{O}] \sim
\mathcal{Q}_\mathrm{O}$ and then:
\begin{equation}
  \frac{1}{n_{\sigma}}
  \left(\frac{\mathcal{Q}_\mathrm{l}}{\mathcal{Q}_\mathrm{O}} - 1
  \right) >
  \frac{\sigma[\mathcal{Q}_\mathrm{O}]}{\mathcal{Q}_\mathrm{O}}
\end{equation}
If we include the SNR via Eq.~\ref{eq:ratios} and Eq.~\ref{eq:def:snr} 
we can obtain the limit condition for the signal-noise ratios:
\begin{equation}
  \rho^2 \equiv \frac{1}{n^2_{\sigma}}
  \left(\frac{\mathcal{Q}_\mathrm{l}}{\mathcal{Q}_\mathrm{O}} - 1
  \right)^2 = \mathrm{SNR}^{-2}_{\broadband} +
  \mathrm{SNR}^{-2}_{\narrowband}
  \label{eq:optimal}
\end{equation}
The SNR depends on the magnitude of the object, the sky background,
the aperture where the photometry is measured, some parameters of
detector (gain and readout noise) and in the exposure time. In our
problem the only free parameters are the exposure times, as the others
are fixed by the observing conditions and by the particular emission
line object that we wish to detect.

We intend to make the dependence in exposure time explicit.  For
exposures where the readout noise of the detector is negligible, the
signal-noise ratio goes roughly as the square root of the exposure
time when the other free parameters are fixed.  For the broad-band, we
comprise in the constant $\alpha$ all the terms not depending on time:
\begin{equation}
  \mathrm{SNR}_{\broadband} = \alpha \,\sqrt{t_{\broadband}}
  \label{eq:snr:explicit1}
\end{equation}

The signal-noise ratio in both bands is related through the ratio of
exposure times. If we start from Eq.~\ref{eq:def:snr}, we can easily
obtain (without any approximation):
\begin{equation}
  \left(\frac{\mathrm{SNR}_{\narrowband}}{\mathrm{SNR}_{\broadband}}\right)^2
  =
  b\,\mathcal{Q}^2\,\left(\frac{t_{\narrowband}}{t_{\broadband}}\right)
  \left(\frac{\left(1 + q_{\broadband}\right) + \frac{A G
  R^2}{\dot{N}_{\broadband} t_{\broadband}}}
  {\left(\mathcal{Q}+q_{\broadband} \, \mathcal{Q}_\mathrm{S}\right) +
  \frac{A G R^2}{b \dot{N}_{\broadband} t_{\narrowband}}}\right)
\end{equation}

For long exposures in both bands, the terms including $R^2$ can be
neglected and we obtain simply:
\begin{equation}
  \left(\frac{\mathrm{SNR}_{\narrowband}}{\mathrm{SNR}_{\broadband}}\right)^2
  =
  b\,\mathcal{Q}^2\,\left(\frac{t_{\narrowband}}{t_{\broadband}}\right)
  \left(\frac{1 + q_{\broadband}}
  {\mathcal{Q}+q_{\broadband}\,\mathcal{Q}_\mathrm{S}}\right) =
  \beta^2 \left(\frac{t_{\narrowband}}{t_{\broadband}}\right)
       \label{eq:snr:ratio}
\end{equation}
being $\beta$ the following:
\begin{equation}
  \beta^2 = b\,\mathcal{Q}^2\,\left(\frac{1 + q_{\broadband}}
       {\mathcal{Q}+q_{\broadband}\,\mathcal{Q}_\mathrm{S}}\right)
       \label{eq:snr:beta}
\end{equation}

The parameter $\beta$ has limit cases depending on $q_{\broadband}$.
If $q_{\broadband} \to 0$, the noise is dominated by the Poisson noise
of the source flux:
\begin{equation}
  \beta^2 = b\,\mathcal{Q}
\end{equation}
If $q_{\broadband} \to \infty$, the
equation is dominated by the Poisson noise of the sky
background. This is the case for faint objects in the images:
\begin{equation}
  \beta^2 = b\,\frac{\mathcal{Q}^2}{\mathcal{Q}_\mathrm{S}}
\end{equation}

We also make explicit the dependence on exposure time of the
narrow-band combining Eq.~\ref{eq:snr:explicit1} and
Eq.~\ref{eq:snr:ratio} to obtain:
\begin{equation}
  \mathrm{SNR}_{\narrowband} = \alpha \,\beta \,\sqrt{t_{\narrowband}}
  \label{eq:snr:explicit2}
\end{equation}

Equation~\ref{eq:optimal} can be written:
\begin{equation}
  \rho^2 = \frac{1}{\alpha^2\,t_{\broadband}} +
  \frac{1}{\alpha^2\,\beta^2\,t_{\narrowband}}
  \label{eq:optimal2}
\end{equation}

\begin{figure}
    \plotone{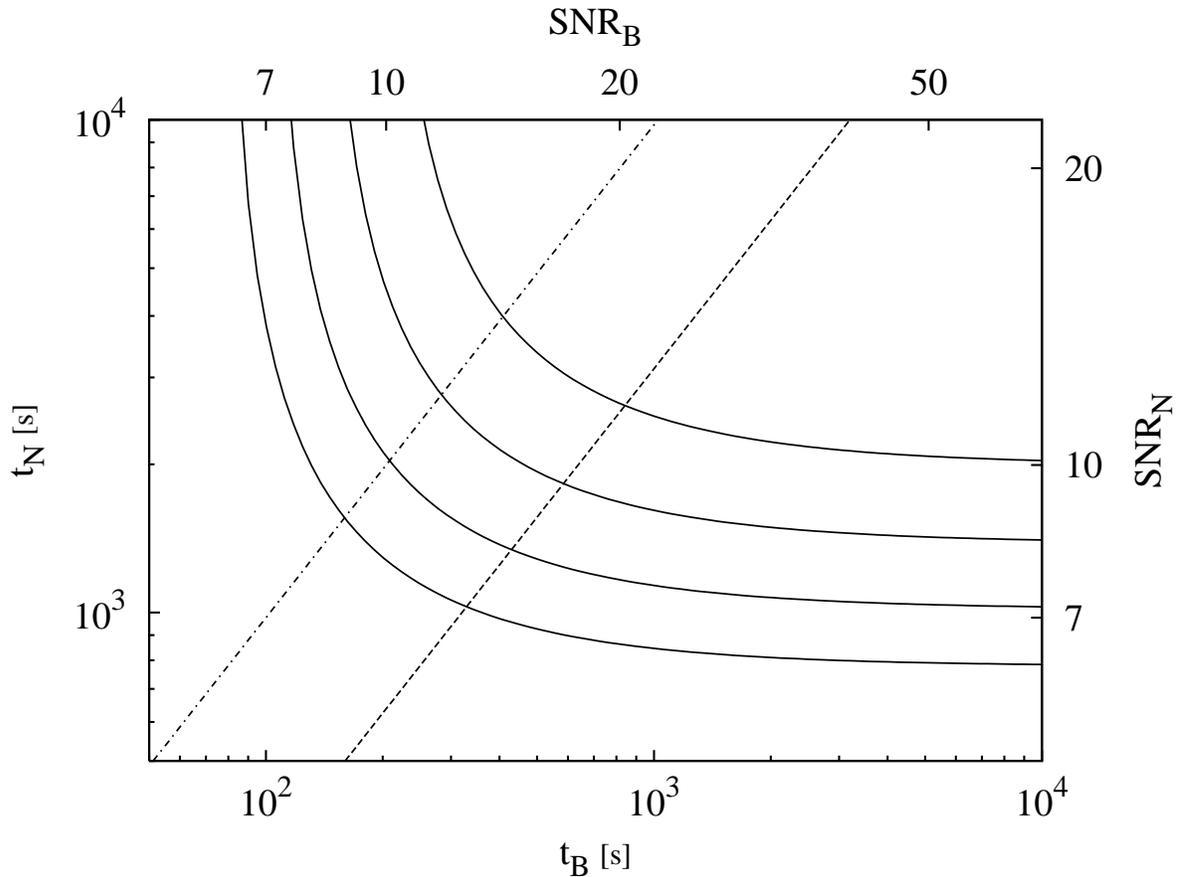}
    \caption{Exposure times in broad- ($t_{\broadband}$) and
      narrow-bands ($t_{\narrowband}$) that allow the detection of an
      object with a given color ratio $\mathcal{Q}_\mathrm{l}$.  The
      corresponding SNR is also represented. The different curves in
      solid line correspond to different values of the parameter
      $\rho$ (see text for the definition).  From left to right
      $\rho=0.16$ (this value is used in examples in text), 0.14, 0.12
      and 0.10.  The dashed line represents the condition of minimum
      total exposure times and the dash-dotted line the condition of
      equal SNR in both bands.  }
    \label{fig:optimal}
\end{figure}

We try to detect 5-$\sigma$ over the mean objects with
$m_{\narrowband}$ = 22 and $\mathcal{Q}_\mathrm{l}$ = 1.8 (this
represents a color excess of 0.6 magnitudes ).  Assuming a mean color
of the objects without emission line of $\mathcal{Q_\mathrm{O}}=1$, we
immediately compute $\rho=0.16$.

With these filters, the color of the sky background is
$\mathcal{Q}_\mathrm{S}=0.23$ and $b=0.024$.  From the exposure time
calculator of the INT WFC,
SIGNAL\footnote{http://www.ing.iac.es/ds/signal/}, we obtain that, for
an object of $m_{\broadband}$ = 22, with seeing of 1\arcsec{} and a
brightness of the sky background of 19.60 mag/arcs$^2$ (typical for a
dark night), the SNR in the broad band is the square root of the
exposure time times $\alpha$ = 0.7.

For the narrow band we compute $\beta$. In this case,
$q_{\broadband}$, computed from the magnitude of the sky background
and the magnitude of the objects without emission line, is $\sim$
300. This means that the noise is dominated by the contribution of the
sky background and $\beta$ = 0.32.

In Figure~\ref{fig:optimal} we represent the exposure times in both
bands that fulfill the condition in Eq.~\ref{eq:optimal2} for
different values of $\rho$, from 0.16 (the example we are considering)
to 0.10 in steps of 0.02.  For each value there is a set of different
exposure times in both bands (given by the curves) that fulfill the
conditions.

Each curve is limited by two asymptotic limits, obtained from the
condition of infinite SNR in one of the bands:
\begin{eqnarray}
  \mathrm{For} \ t_{\narrowband} \to \infty \ &
  t^{\mathrm{lim}}_{\broadband} & = \frac{1}{(\rho\,\alpha)^2}\\
  \mathrm{For} \ t_{\broadband} \to \infty \ &
  t^{\mathrm{lim}}_{\narrowband} & = \frac{1}{(\rho\,\alpha\,\beta)^2}
\end{eqnarray}

In the current example, the limit exposures are
$t^{\mathrm{lim}}_{\broadband}$ = 80 s and
$t^{\mathrm{lim}}_{\narrowband}$ = 780 s. As the curves lay over their
asymptotes, this exposure times are lower limits.

The problem of selecting the optimal exposure times in order to
optimize selection of objects with the narrow-band filter still exists
at this point.  We have to impose one additional condition to extract
a pair $t_{\broadband}$, $t_{\narrowband}$ from Fig~\ref{fig:optimal}.
It seems natural to obtain the point that minimizes the total exposure
time $t_{\broadband} + t_{\narrowband}$. This condition is equivalent
to minimize the sum of inverse squared SNRs with the constraint of
constant $t_{\broadband} + t_{\narrowband}$.

The exposure times are:
\begin{eqnarray}
  t_{\broadband} & = & \frac{\beta + 1}{\rho^2\,\alpha^2\,\beta}\\
  t_{\narrowband} & = & \frac{\beta + 1}{(\rho\,\alpha\,\beta)^2} =
  \frac{1}{\beta}\,t_{\broadband}
\end{eqnarray}

In terms of the signal-noise ratio, this last equation can be written
as:
\begin{eqnarray}
  \mathrm{SNR}^2_{\broadband} & = & \frac{\beta + 1}{\beta \,\rho^2}\\
  \mathrm{SNR}^2_{\narrowband} & = & \frac{\beta + 1}{\rho^2} =
  \beta\, \mathrm{SNR}^2_{\broadband}
\end{eqnarray}
      
For our particular example, the exposure times are: $t_{\broadband}$ =
330 s and $t_{\narrowband}$ = 1030 s.  The corresponding signal to
noise ratios are $\mathrm{SNR}_{\broadband}$ = 13 and
$\mathrm{SNR}_{\narrowband}$ = 7.

From the previous equations we can extract an important conclusion.
As typically $\beta < 1$, the exposure times in the narrow band are
greater than the exposure times in the broad band, but the SNR in the
narrow band is lower than the SNR in the broad band. This can be a
problem in some extreme cases.  The narrow-band image is used normally
to detect the objects. Hence, it is necessary to have enough
signal-noise ratio in the narrow-band image to be able to detect the
objects in it. Note that these exposure times do not account for the
overheads proper of observations (dithering, readout time of the
detector, etc.).

Other conditions can be applied to Eq.~\ref{eq:optimal2} and different
exposure times are obtained. If we wish to assure that the SNR is high
simultaneously in both images, we can impose the condition of equal
SNR to the objects without emission line. Both terms on the right hand
side of Eq.~\ref{eq:optimal2} are equal.  The exposure times are:
\begin{eqnarray}
  t_{\broadband} & = & \frac{2}{\rho^2\,\alpha^2}\\
  t_{\narrowband}  & = & \frac{2}{(\rho\,\alpha\,\beta)^2} =
  \frac{1}{\beta^2}\,t_{\broadband}
\end{eqnarray}
and the common signal-noise ratio is:
\begin{equation}
  \mathrm{SNR}^2_{\broadband} = \mathrm{SNR}^2_{\narrowband}= \frac{2}{\rho^2}
\end{equation}

If we apply the condition of equal SNR to our example, we obtain
$t_{\broadband}$ = 160 s and $t_{\narrowband}$ = 1560 s, whereas the
SNR in both bands is $\mathrm{SNR}_{\broadband}$ =
$\mathrm{SNR}_{\narrowband}$ = 9. The total time in this case is 1720
s, that can be compared with the total exposure time of 1360 s in the
optimal case. We have represented in Fig.~\ref{fig:optimal} the lines that
correspond to minimum exposure time (dashed) and equal SNR
(dash-dotted).

Other possibility is to work in the asymptote of long broad-band
exposure times. With the availability of multicolor imaging surveys,
deep broad-band public images can be used instead of observing
simultaneously in different bands, including the narrow-band.  In this
case, the flux limits are completely determined by the depth of the
narrow-band image (in Eq.~\ref{eq:optimal},
$\mathrm{SNR}^{-2}_{\broadband} \ll \mathrm{SNR}^{-2}_{\narrowband}$).

As we have shown, the condition of minimum exposure times provides
high enough SNRs to carry out object detection in the narrow-band
image. In the SNR obtained is not enough, other conditions can be
imposed. The presented exposure times are only lower limits, obtained
from the limit condition of Eq. \ref{eq:optimal}.  Longer exposure
times, if possible, provide better SNRs in both bands that allow
determining the line flux and EW of the emission line with lower
uncertainties. 

The goals of the emission-line survey should be considered when
deciding exposure times. Searches for relatively rare objects should
survey relatively more area at the cost of getting lower SNR on the
narrow-band filters. On the other side, very deep narrow-band survey
can select candidates that will have very difficult follow up
spectroscopy. In this last case, very deep narrow-band imaging is
appropriated, in order to recover line fluxes with very low
uncertainty.

\section{Summary and Conclusions}
\label{sec:conclusiones}
The main results of this work are as follow:

\begin{enumerate}

\item We discuss the contamination of a narrow-band survey, either by
  galaxies or stars and show different classification schemes that can
  be used to separate their different contributions. In general, multicolor
  photometry can be used to classify stars and ELGS at different redshifts.
  Morphological criteria can also help in the classification.
  
\item For different narrow-band filters, we compute the mean colors of
  stars and galaxies.  With all the sample filters, the galaxies with
  emission lines exhibit a color excess and they can be selected as
  noted by previous authors.  Stars of different types can be selected
  depending on the shapes of the filters involved. A particular
  analysis is needed in each case. In general, late type stars have
  molecular bands in the regions studied that can be easily confused
  with emission lines.

\item The line flux and equivalent width are obtained with different
  assumptions about the continuum and different filter layouts.  line
  We also confirm that the assumption of an infinitely thin line is
  appropriate for most physical cases.

\item We study the case of several lines entering simultaneously in
  the narrow-band filter. This problem arises for \HA{} plus \NII{}
  and \OIII{} plus \HB. In each case we compute the correction due to
  the contribution of the other lines. For very narrow filters, using
  common line ratios can underestimate the line flux, as the
  additional lines do not enter in the filter.
    
\item We study the magnitude-color diagram used to select the
  candidates and we compute an analytic form of the standard deviation
  of the color of the objects. The standard deviation depends on a
  number of parameters, namely the exposure times in both bands, the
  colors of the object and of the background, the surface brightness
  of the sky and the magnitude of the object.  The equation of
  $\sigma[\mathcal{Q}]$ can help when there are not enough objects
  detected to calculate the standard deviation from the data
  available.
    
\item We show how to compute the redshift range where a candidate can
  be included in the sample. This range is used to compute the
  comoving volume associated with each object. Using a global comoving
  volume for all the object tends to under estimate the volume for
  faint objects and objects selected near the selection curve.

\item We provide equations to compute the optimal exposure time in the
  narrow band given the exposure time in the broad band. Given a color
  ratio for an object, we show the equation that relates the exposure
  times in both bands needed to effectively select the object.

\item Additionally, if we impose the condition of minimum total
  exposure time, we obtain directly the exposure times in both
  bands. These equations are invaluable tools when planing narrow-band
  observations, as exposure time calculators usually only manage
  broad-band filters.
\end{enumerate}

\acknowledgments 
This research was supported by the Spanish
\emph{Programa Nacional de Astronom\'{\i}a y Astrof\'{\i}sica} under
grant AYA2003-1676.

The authors would like to thank the referee, David Thompson, for 
insightful comments that improved the paper.

{\it Facilities:} \facility{CAO:2.2m}, \facility{ING:Newton }.
\bibliographystyle{apj}
\bibliography{referencias,astroph,local}
\end{document}